\documentclass[sigconf,nonacm]{acmart}
\AtBeginDocument{%
  }

\usepackage{enumitem,subcaption,multirow,tabularx}
\usepackage{threeparttable}
\usepackage{booktabs,array,longtable}

\usepackage{tikz}

\definecolor{deployLowBg}{HTML}{E4F2EC}
\definecolor{deployLowFg}{HTML}{196343}
\definecolor{deployModerateBg}{HTML}{FFF0D9}
\definecolor{deployModerateFg}{HTML}{865008}
\definecolor{deployHighBg}{HTML}{F9E4E6}
\definecolor{deployHighFg}{HTML}{9B2936}
\providecommand{\deploylow}{\begingroup\setlength{\fboxsep}{2pt}\colorbox{deployLowBg}{\textcolor{deployLowFg}{\strut\textbf{Low}}}\endgroup}
\providecommand{\deploymoderate}{\begingroup\setlength{\fboxsep}{2pt}\colorbox{deployModerateBg}{\textcolor{deployModerateFg}{\strut\textbf{Moderate}}}\endgroup}
\providecommand{\deployhigh}{\begingroup\setlength{\fboxsep}{2pt}\colorbox{deployHighBg}{\textcolor{deployHighFg}{\strut\textbf{High}}}\endgroup}

\providecommand{\defreq}[2]{%
  \mbox{\csname deficon#1\endcsname\,\textbf{#2}}}
  
\begin{document}

\title{Can We Stop The Ads? Taxonomy and Characterization of Smartphone Splash Ads and Existing Countermeasures}

\author{%
Shuhao Zhang\textsuperscript{1},
Xinyu Liu\textsuperscript{1},
Ziyu Shao\textsuperscript{1},
Yuqing Yang\textsuperscript{2},
Yan Long\textsuperscript{1}}

\affiliation[obeypunctuation=true]{%
  \institution{%
    \textsuperscript{1}The Hong Kong University of Science
    and Technology (Guangzhou)}, \country{China}\\
  \institution{%
    \textsuperscript{2}Macau University of Science and Technology},
    \country{China}%
}

\email{{szhang515,zshao787}@connect.hkust-gz.edu.cn}
\email{xinyuliu@hkust-gz.edu.cn; yuqingyang@must.edu.mo}
\email{yanlong@hkust-gz.edu.cn}

\renewcommand{\shortauthors}{S. Zhang et al.}
\settopmatter{authorsperrow=1}

\begin{abstract}

Splash ads are full-screen advertisements that pop up and appear as the first interaction page when users start an app, often tricking users into unknowingly activating certain trigger mechanisms, such as moving the phone to redirect users to other profit-driven third parties. 
So far, splash ads have already caused significant real-world impacts, ranging from significantly delaying emergency response to distracting drivers, as well as degrading accessibility of apps to vision-impaired users. We analyze 108 documented implementations\footnote{Inventory: \url{https://github.com/SENSE-Lab-Security/SplashAds-Countermeasures}} of advertising defenses to examine their applicability to splash ads and the requirements users face when deploying them.

Our analysis identifies substantial deployment barriers, including device rooting or jailbreaking, runtime code injection, and application modification. Options without these requirements can still involve additional permissions, rule maintenance, source compilation, or payment.
In our evaluation of 13 configurations of 11 tools across 10 popular apps, only one tool prevented the target ad-triggered navigation across all ten apps. It required Accessibility permission, and ads remained visible for approximately one second before dismissal. Other tested configurations failed to prevent navigation or, in some cases, left host apps unable to launch or stuck on the ad page.
We further analyze the outstanding challenges and possible future directions, highlighting the urgent need to incentivize smartphone manufacturers to provide more friendly and regulated platforms.

\end{abstract}

\maketitle

\pagestyle{plain}
\thispagestyle{plain}

\section{Introduction}

Splash screen advertisements, or Splash Ads, are a common form of advertising. These ads are displayed when users launch a mobile application, before they can access its main functions. While this business model ensures that users see advertisements, it can also cause problems beyond inconvenience. Recent reports describe older adults and people with visual impairments being misled into clicking advertisements when they intended to use the application~\cite{lu2026blindads,cmg2026elderly}. There are even reports of victims contacting the fire department being forced to watch an advertisement of more than half a minute before being able to upload the required video ~\cite{yang2026fire}.

Unfortunately, users have limited approaches to fight these ads. First, blocking Splash Ads can be technically challenging. The advertisements may be embedded in the application or downloaded from advertising servers, and users have little control over either process. Second, users may struggle to have their complaints addressed. Vendors provide limited feedback channels and may have little incentive to remove advertisements that generate revenue.

Third, while national-level policies and regulations have already been announced, technical compliance by the main stakeholders still falls short.  To understand how users can better address these problems, we need to take the perspective of ordinary, non-expert users, and examine how Splash Ads work, what existing tools can do, and what prevents users from using these tools effectively.

As such, in this study, we investigate the landscape of Splash Ads in mobile ecosystems including Android and iOS particularly aiming to answer four research questions:

\begin{enumerate}
    \item What mechanisms do Splash Ads use, and how can these ads be classified?
    \item Does platform difference affect the adoption and triggers of Splash Ad?
    \item What mitigation methods are available, and what technical challenges do users face when deploying them?
    \item What changes could help users better avoid or reduce the negative effects of Splash Ads?
\end{enumerate}

To answer these questions, we first conduct a comprehensive literature review of news report, technical documents, and GitHub repositories to collect incidents around Splash Ads, mechanisms behind Splash Ads, and mitigation tools from individual developers. Then, we summarize our findings to comprise a taxonomy of available approaches to enable Splash Ads. We find that Splash Ads generally rely on triggers to initiate, and such triggers can mainly be categorized into four types: single tapping, screen swiping, shaking the device, and device rotation. We also observe Splash Ads in over 30\% of the top 30 Android applications we examined, with some applications implementing multiple trigger mechanisms.

In terms of evaluating mitigation approaches, we perform a differential and empirical analysis, testing collected Splash Ad mitigation tools to evaluate their effectiveness and ease to use. Our investigation reveal that existing defenses are difficult for ordinary users to deploy and provide inconsistent protection. Among the 108 countermeasure instances we study, 39 require rooting or jailbreaking, runtime code injection, or APK modification. Even tools without these requirements may require additional permissions, rules, source compilation, or other configuration. In our empirical evaluation, only 28 of 130 tool-app combinations successfully prevented the target ad-triggered navigation, and some defenses even prevented host applications from launching or left them stuck on the ad page.

With these, we summarize the technical challenge around Splash Ads, namely why it is so challenging to stop Splash Ads, despite that there are numerous mitigation tools. These challenges include the difficulty of separating advertising behavior from normal application functionality, dependence on rules and configurations that may be invalidated with software updates, extensive deployment and runtime requirements, and potential legal conflicts around Splash Ad blocking tools. In the end, we propose future work directions to tackle these challenges and highlight the need of collective effort in mitigating hazards of Splash Ads in emergency or usability-centric environments.

In short, this paper makes the following contributions:

\begin{itemize}
    \item We provide a systematic taxonomy of Splash Advertisements based on their mechanisms.
    \item We perform comprehensive analysis of the ecosystem around Splash Ads, including the mitigation tools developed to fight against abusive Splash Ads.
    \item We perform a large empirical evaluation on the effectiveness and usability of Splash Ads mitigation tools.
    \item We propose approaches to mitigate Splash Ads issues in hazardous and impaired environments.
    \item We provide a public inventory of the surveyed tools, with links to their sources and documentation of their deployment requirements and limitations.
\end{itemize}

\section{Background \& Motivation}

This section introduces the real-world problems and relevant research results that motivate our work.  

\subsection{Resulting Real-World Incidents}
\label{sec:realworld}

Recent incidents reported by Chinese news media illustrate that intrusive
mobile app advertisements can create consequences beyond ordinary inconvenience. Below we highlight several outstanding concerning examples. 

\textbf{Threatening Driving Safety.}
In June 2026, a driver reported that he opened a navigation application through
voice control while driving, without touching the phone. Vehicle vibration
subsequently triggered a high-sensitivity ``shake-to-open'' splash
advertisement, causing the interface to redirect to an e-commerce application
and distracting him while driving~\cite{du2026navigation}. \textbf{\textit{The driver reported
that the incident nearly resulted in a collision.}} In a follow-up test, reporters
also confirmed that two mainstream navigation applications displayed
five-second splash ads for which shaking the phone could open the
advertisement details.

\textbf{Hindering Emergency Response.}
In September 2026, a Shenzhen resident called the emergency service after
observing a fire in a neighboring residence and was asked to upload a video
through a link sent by the dispatcher. Opening the link launched the phone
browser app, whose splash ad appeared before the upload page.  While
attempting to dismiss the ad, the user accidentally closed the
browser and had to reopen the link, \textbf{\textit{delaying the urgent video upload by approximately
30--60 seconds}}. The local fire department subsequently
clarified that the advertisement originated from the user's browser app rather than
the emergency-service link~\cite{yang2026fire}.

\textbf{Causing Accessibility Barriers.}
Intrusive advertisements can impose particularly severe barriersB on users who
rely on accessibility services. A September 2026 report described viral footage
of a \textbf{\textit{visually impaired user}} in Jiangxi repeatedly attempting to operate
a smartphone while advertisement pop-ups continued to appear and redirect the
interface~\cite{lu2026blindads}. Phoenix Technology subsequently tested several
applications using Android screen-reader mode and observed that important
advertisement controls, including close buttons, could lack meaningful
accessible labels, making it difficult for visually impaired users to
distinguish dismissal controls from advertisement content.

\begin{figure}[t]
  \centering
  \includegraphics[width=0.95\columnwidth]
  {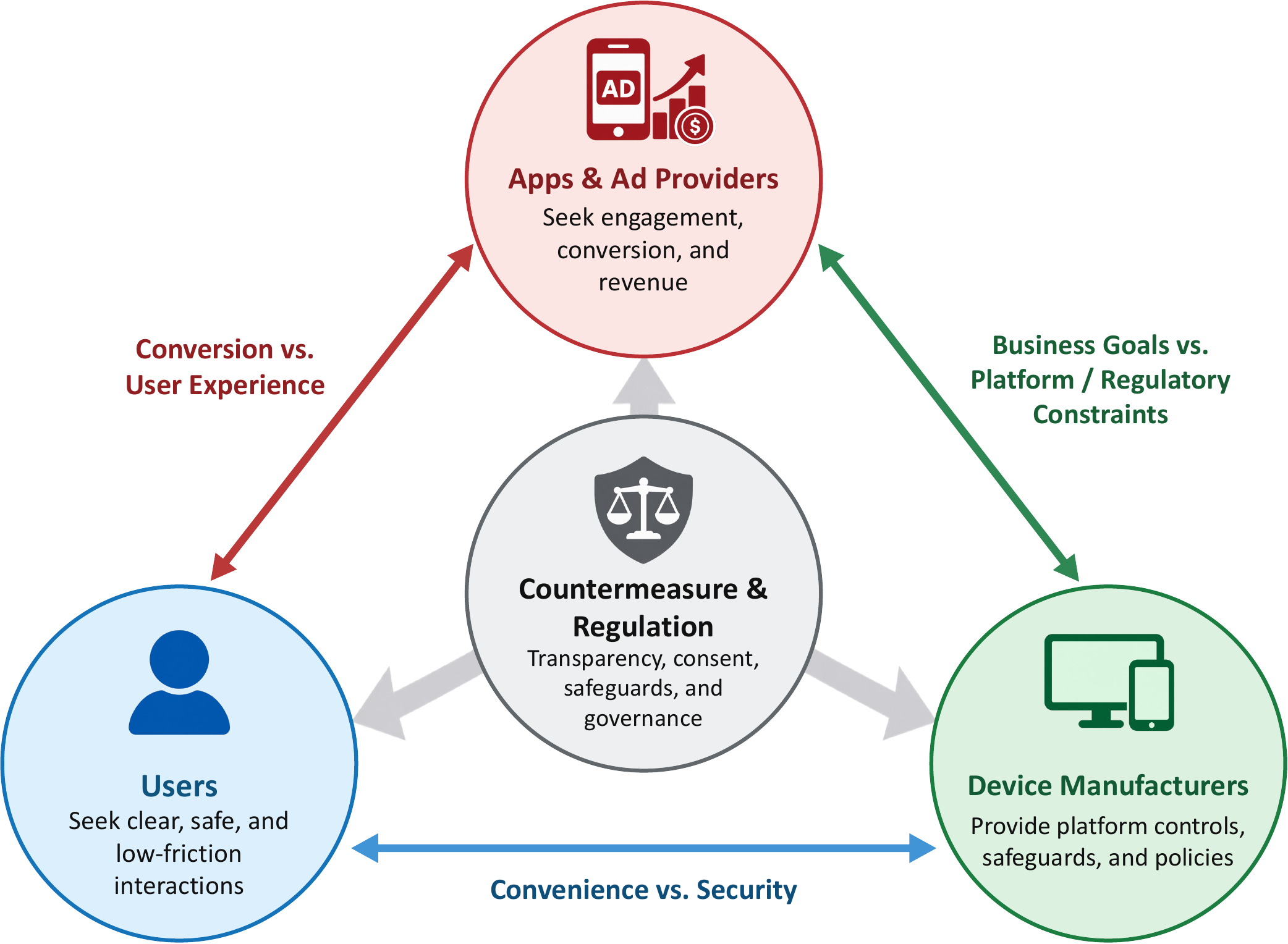}
  \caption{Competing objectives between the stakeholders in mobile splash advertising. Possible countermeasures and regulations may help improve balance between these parties.}
  \label{fig:tension}
\end{figure}

\textbf{Impact on Elderly Users.}
In another widely reported incident, an elderly user in Huangshan attempted to
take a photograph but encountered \textit{\textbf{more than 20 advertisement pop-ups or
redirects within approximately 30 seconds and eventually gave up the
task}}~\cite{cmg2026elderly}. The same investigation reported more than 2,800
complaints related to pop-up advertisements and documented complaints involving
small close buttons, high-sensitivity ``shake-to-open'' triggers, and unwanted
application redirection.

The examples above are far from a comprehensive list of the negative real-world impact of smartphone splash ads. Together, these incidents show that aggressive advertisement interactions can
affect application
availability and even physical safety of users. More fundamentally, they demonstrate a significant and unaddressed tension (Figure~\ref{fig:tension}):
\textbf{\textit{user actions or environmental inputs that do not clearly express an intention to
open an advertisement can nevertheless be exploited and misinterpreted to generate profit-driven
redirection operations, and the process is largely out of users' controls}}.

\subsection{User Attitudes and Demand for Control}
\label{sec:user_demand}

Available evidence shows not only widespread dissatisfaction with intrusive
splash advertisements, but also a concrete demand for mechanisms that give
users greater control over advertisement activation and redirection. 

A poll conducted by the Jiangsu Consumer Protection Committee shows that 78\% of
respondents reported frequently encountering ``shake-to-open'' advertisements,
92\% expressed dislike toward this form of splash advertising, and 90\%
considered the behavior an infringement of their rights. More importantly,
76\% supported stopping shake-triggered splash advertisements or related
behaviors. 

Empirical research focusing specifically on splash advertisements reaches a
similar conclusion regarding the importance of user control. Zheng et
al.~\cite{zheng2025splash} surveyed users in China and found that the
responsiveness and visual salience of the skip option increase users'
\emph{perceived control}, which in turn improves their attitudes toward both
the splash advertisement and the hosting application. Complementary controlled
experiments by An et al.~\cite{an2026splash} found that the presence of splash
advertisements decreases users' attitudes toward advertisements and their
satisfaction with the corresponding applications, largely because such
advertisements are perceived as obstacles to the users' intended goals.
These findings suggest that an important user requirement is
effective and readily accessible control over whether an advertisement is
viewed, dismissed, or allowed to initiate a navigation action. Unfortunately, as of September 2026, splash advertisement SDKs and apps using these services still do not provide such controls. 

\begin{table}[t]
\centering
\caption{Semi-drive-by splash-ad prevalence and trigger distribution reported by Wu et al.~\cite{wu2026adhive}.}
\label{tab:semi_drive_by}
\small
\setlength{\tabcolsep}{3.5pt}
\begin{tabular}{lrrcc}
\toprule
\textbf{Market} &
\textbf{Tested} &
\textbf{Detected} &
\textbf{Touch} &
\textbf{Motion} \\
\midrule
VIVO
& 5,262
& 261 (4.96\%)
& 12 (4.6\%)
& 249 (95.4\%) \\

Xiaomi
& 5,998
& 128 (2.13\%)
& 81 (63.3\%)
& 47 (36.7\%) \\

Wandoujia
& 11,682
& 198 (1.70\%)
& 18 (9.1\%)
& 180 (90.9\%) \\

YingYongBao
& 8,881
& 191 (2.15\%)
& 13 (6.8\%)
& 168 (88.0\%) \\
\midrule
\textbf{Total}
& \textbf{31,823}
& \textbf{778 (2.44\%)}
& \textbf{124 (15.9\%)}
& \textbf{654 (84.1\%)} \\
\bottomrule
\end{tabular}
\end{table}

\subsection{Policy Expectations and Current Status}
\label{sec:policy}

The current policy and industry framework in China already establishes a
relatively clear expectation for more user-friendly splash advertising, but the actual executive compliance of existing apps still shows a significant gap. 

Prior work
notes that splash advertisements have been subject to dedicated rectification
and that users should retain effective control, such as readily accessible
skip options~\cite{an2026splash}. For motion-triggered advertisements, recent
technical work further adopts established industry thresholds specifically
designed to prevent ordinary activities---such as walking, picking up a phone,
or riding in a vehicle---from being interpreted as intentional advertisement
engagement~\cite{wu2026adhive}. In the intended compliant case, advertisement
navigation should therefore follow a clear and deliberate user action, while
incidental touch or device motion should not cause redirection.

The remaining challenge lies largely in how these requirements are implemented
by applications, advertising SDKs, and advertising networks. Empirical studies
continue to observe substantial use of deceptive or overly sensitive
interaction mechanisms in deployed apps. Long et al.\ found dark UI patterns
in 82\% of 150 popular Chinese apps, including misleading advertisement
interactions and easy-to-trigger behaviors~\cite{long2023darkui,long2025armour}.
More recently, Shang et al.\ found shake-to-open advertisements to be
particularly prevalent in their Chinese app dataset~\cite{shang2025adsdp}. Wu et al.~\cite{wu2026adhive} examined 31,823 Android applications across four major Chinese app markets and identified 261, 128, 198, and 191 applications exhibiting such behavior in VIVO, Xiaomi, Wandoujia, and YingYongBao, respectively. Among these detected cases, sensor-triggered mechanisms accounted for 95.4\%, 36.7\%, 90.9\%, and 88.0\%, respectively, highlighting the importance of motion-sensor-based activation in semi-drive-by advertising.

\textbf{Gap Between Policy and Deployment.} These findings suggest that the main practical
gap is no longer the absence of well-defined expectations, but uneven
compliance and enforcement within the advertising ecosystem. This motivates
our investigation of technical countermeasures that end users may deploy
immediately to protect themselves, without requiring every application or
advertising provider to first adopt the intended behavior.

\section{Taxonomy of Intrusive Splash Ads}
\label{sec:pipeline}

\begin{figure}[t]
    \centering
    \includegraphics[width=\columnwidth]{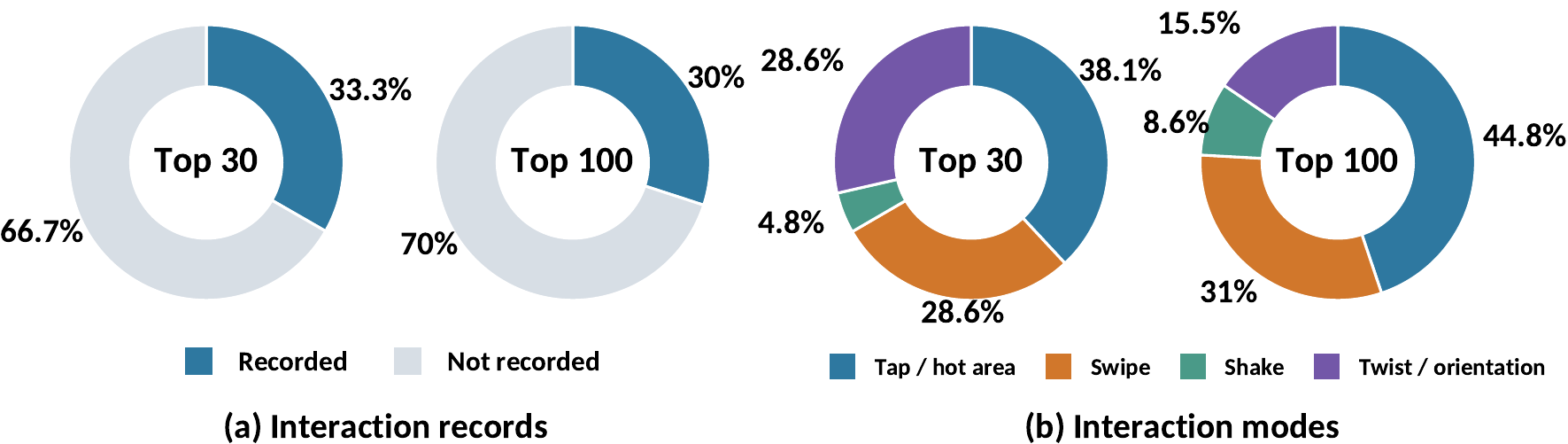}
    \caption{Interaction records and interaction modes observed in popular Android apps selected from the QuestMobile Top-30 and MoonFox Top-100 rankings for June 2026. All apps were tested in September 2026.}
    \label{fig:interaction-characteristics}
\end{figure}

We further examined the Android versions of the top 30 applications in QuestMobile's 2026 H1 China Mobile Internet APP User Scale Ranking and observed splash advertisements in 10 of them (33.3\%). The result shows that splash advertising remains common among today's most widely used applications, making its interaction mechanisms particularly relevant to a large population of mobile users.Figure~\ref{fig:interaction-characteristics} summarizes the observed occurrence of splash ads and the distribution of trigger types in the Top-30 and Top-100 app samples.

\begin{figure*}[t]
  \centering
  \includegraphics[width=1.99\columnwidth]
  {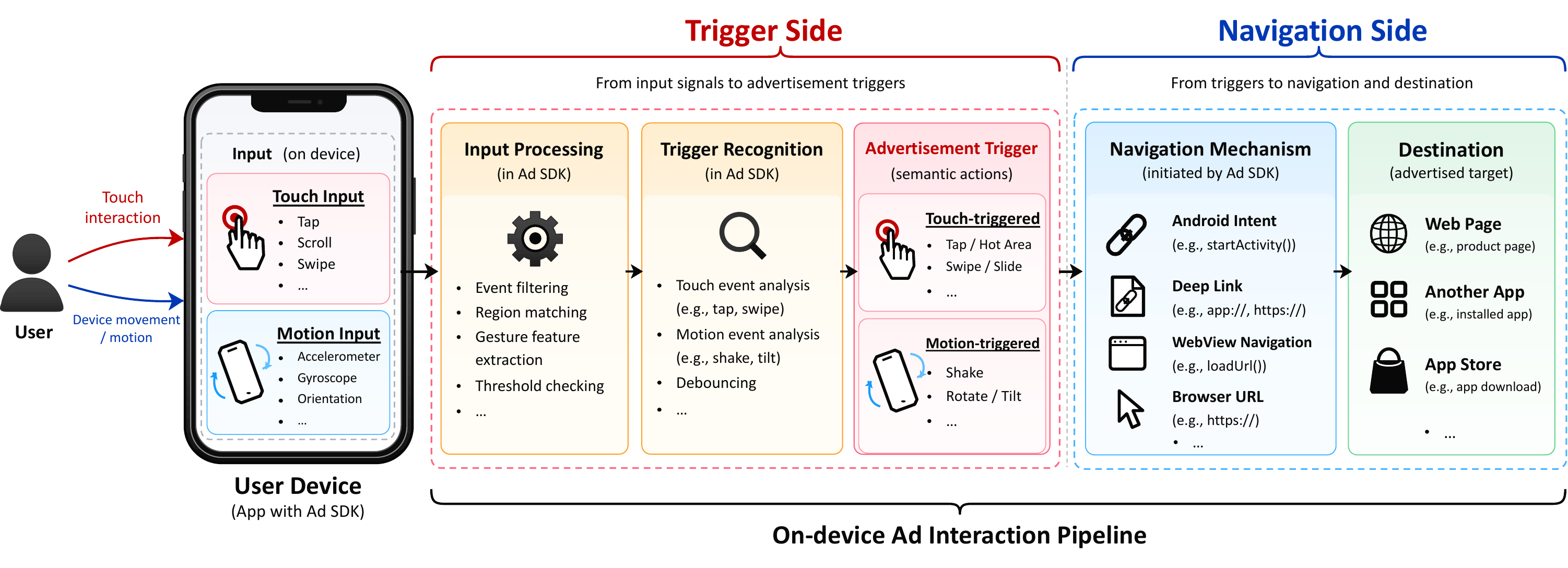}
  \caption{The common pipeline of splash ads.}
  \label{fig:ad-pipeline}
\end{figure*}

\textbf{Trigger-to-Navigation Pipeline.}
Despite their diverse interaction designs, interactive splash advertisements generally follow a common on-device pipeline from user input to advertisement navigation, as illustrated in Fig.~\ref{fig:ad-pipeline}. 

We divide this pipeline into two stages: the \emph{Trigger Side} and the \emph{Navigation Side}. 

The Trigger Side determines \emph{when and how} an advertisement is activated. Based on the source and interpretation of the input, we classify these mechanisms into two major categories: \emph{touch-triggered} advertisements, including \emph{Tap/Hot Area} and \emph{Swipe/Slide}, and \emph{motion-triggered} advertisements, including \emph{Shake} and \emph{Rotate/Tilt}. This distinction is important because the two categories rely on different device inputs and recognition logic, and therefore expose different opportunities for intervention.

Once an advertisement has been triggered, the Navigation Side determines \emph{where and how} the user is redirected. Although splash advertisements employ heterogeneous trigger mechanisms, their resulting actions often converge on a relatively small set of navigation primitives, such as Android Intents, deep links, WebView navigation, and browser URLs. These primitives can subsequently redirect users to webpages, other applications, or app stores.

\begin{table}[t]
\centering
\caption{Interaction trigger mechanisms observed in the top-30 popular Android applications.}
\label{tab:app_triggers}
\resizebox{\columnwidth}{!}{%
\begin{tabular}{l
    >{\centering\arraybackslash}p{0.75in}
    >{\centering\arraybackslash}p{0.75in}
    >{\centering\arraybackslash}p{0.75in}
    >{\centering\arraybackslash}p{0.75in}}
\toprule
& \multicolumn{2}{c}{\textbf{Touch-Triggered}}
& \multicolumn{2}{c}{\textbf{Motion-Triggered}} \\
\cmidrule(lr){2-3} \cmidrule(lr){4-5}
\textbf{Host App}
& \textbf{Tap / Hot Area}
& \textbf{Swipe}
& \textbf{Shake}
& \textbf{Twist / Orientation} \\
\midrule
Baidu          & $\checkmark$ & --           & --           & --           \\
Sogou Input   & $\checkmark$ & $\checkmark$ & --           & $\checkmark$ \\
Baidu Maps    & $\checkmark$ & $\checkmark$ & --           & $\checkmark$ \\
Baidu Input   & $\checkmark$ & $\checkmark$ & --           & --           \\
Weibo         & $\checkmark$ & $\checkmark$ & --           & --           \\
QQ Browser    & $\checkmark$ & $\checkmark$ & --           & --           \\
Tencent Video & --           & --           & --           & $\checkmark$ \\
iQIYI         & --           & --           & --           & $\checkmark$ \\
Tencent News  & $\checkmark$ & $\checkmark$ & --           & $\checkmark$ \\
Mango TV      & $\checkmark$ & --           & $\checkmark$ & $\checkmark$ \\
\bottomrule
\end{tabular}%
}
\end{table}

\subsection{Deceptive Trigger Mechanism}
The feature of intrusive splash ads that users complain about the most is that after they pop up, there will be certain kinds of user action trigger that users often do unwittingly, without users knowing the ads are exploiting this common unintentional behavior as a trigger. Even worse, some of the triggers cannot even be controlled by users. We summarize the common types of triggers below. 

\subsubsection{Tap- and Hot-Area-Triggered Advertisements}

Both tap- and swipe-triggered advertisements use touchscreen input, but differ in their activation conditions. Tap-triggered advertisements initiate navigation when the user taps within a clickable region, without requiring a directional sliding gesture. ``Hot area'' refers to the extent of that clickable region rather than a separate gesture type. Swipe-triggered advertisements instead require finger movement that satisfies conditions such as a minimum displacement or a specified direction. The main usability problem arises when the effective clickable region is substantially larger than the visually indicated advertisement element, including designs in which most or all of the splash screen acts as an advertisement target.
Misleading or deceptive close controls further blur the distinction between advertisement activation and advertisement dismissal because a touch intended to close an advertisement may instead be interpreted as an advertisement click.

\begin{figure}[t]
    \centering
    \includegraphics[width=0.5\columnwidth]{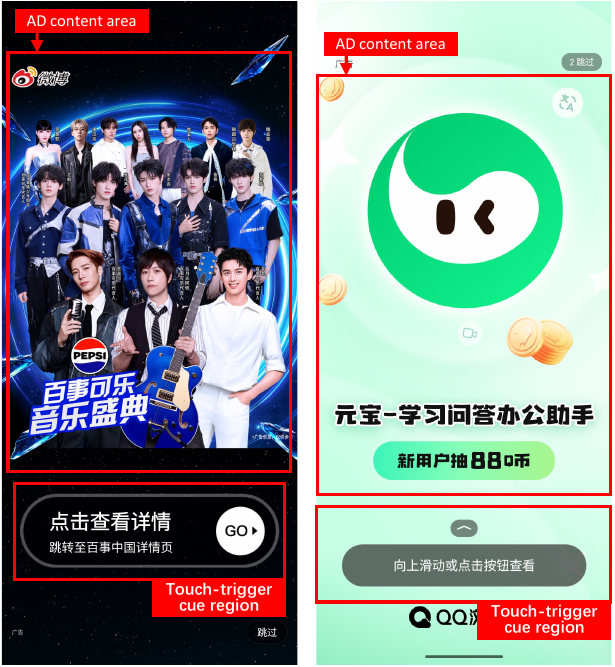}
    \caption{Touch-triggered splash ads with visually separated interaction regions and explicit interaction animations, which guide users toward deliberate interaction with the ad.}
    \label{fig:touch_splash_ads}
\end{figure}

\subsubsection{Swipe-Triggered Advertisements}

Swipe-triggered advertisements infer interaction intent from a sequence of touchscreen coordinates, typically by measuring displacement, direction, duration, velocity, or similarity to a predefined trajectory. A swipe trigger can therefore be activated without access to motion sensors, which makes sensor-permission controls ineffective against this class of advertisements. The technical ambiguity of swipe-based activation comes from the fact that short scrolling gestures, one-handed repositioning, and deliberate advertisement-opening gestures may generate partially overlapping touchscreen trajectories.



\begin{figure}[t]
    \centering
    \includegraphics[width=0.5\columnwidth]{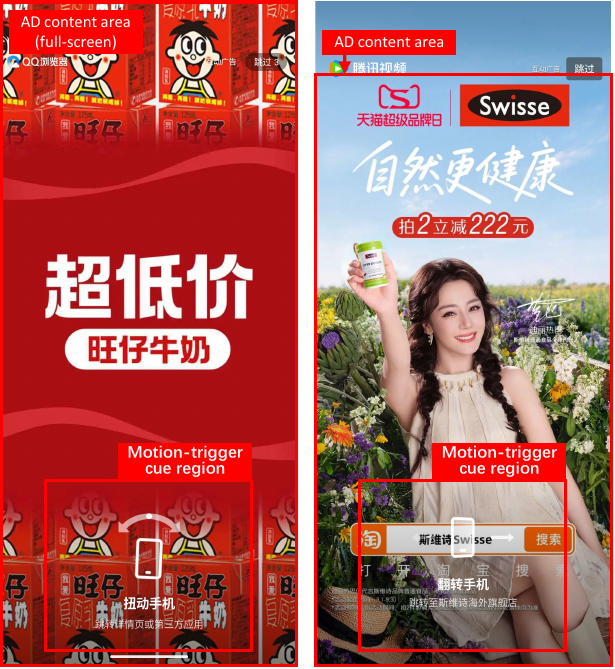}
    \caption{Motion-triggered splash ads with motion instructions overlaid on full-screen video content. The reduced visual salience of these instructions can make the trigger difficult to notice, increasing the risk of unintended advertisement activation.}
    \label{fig:motion_splash_ads}
\end{figure}

\subsubsection{Shake-Triggered Advertisements}
Shake-triggered advertisements monitor accelerometer, gyroscope, or related motion-sensor streams and compare derived motion features against thresholds selected by the application or advertisement SDK. A typical implementation aggregates acceleration magnitude, angular displacement, event duration, or motion velocity and invokes an advertisement callback when one or more thresholds are satisfied. Because natural activities such as picking up the phone, walking, riding in a vehicle, or adjusting device orientation can produce similar sensor patterns, the choice of trigger threshold directly determines the probability of accidental activation. Advertisement SDKs may additionally obtain trigger parameters through remote configuration, allowing motion sensitivity to change without an application update and making one-time application inspection insufficient. By far, shake-triggered ads have been known as the most disturbing and deceptive type of intrusive splash ads.

\subsubsection{Rotation- and Orientation-Triggered Advertisements}
Rotation- or orientation-triggered advertisements use gyroscope, orientation, or fused motion information to detect changes in device pose and trigger navigation after a required angular displacement or motion pattern. These mechanisms are technically similar to shake-triggered advertisements from a permission perspective but can rely on different sensor combinations and feature-extraction logic.
The existence of several motion-triggered interaction styles means that blocking one specific motion pattern does not necessarily remove an advertisement SDK's ability to construct another motion-based trigger.

\subsection{Platform Comparison: Android vs. iOS}

Splash ads exist on both Android and iOS. \textbf{\textit{In our manual observations, splash ads appeared more frequently and consistently on iOS than on Android.}} Among the inspected iOS apps that displayed splash ads, advertisements appeared on nearly every launch, whereas the same Android app did not necessarily display an ad each time it was opened. For example, Baidu Maps displayed splash ads on nearly every observed iOS launch but intermittently or not at all on the tested Android devices. The frequency of advertising interruptions therefore varied across the inspected platforms and devices.

Several factors may explain this difference. The Android and iOS versions of an app may use different display rules, such as showing an ad on every opening or imposing a minimum interval between displays. Advertising platforms provide frequency controls for app-open ads, allowing developers to limit impressions per user over a specified period~\cite{admobFrequencyCapping}. Advertising availability may also differ across devices. Campaign targeting can distinguish operating systems and device models, allowing different devices to receive different advertising opportunities~\cite{googleAdsMobileDeviceTargeting}. The two app versions may also load ads at different times. A preloaded ad can be displayed immediately, whereas an ad requested only when the app opens may not become available before the loading screen ends~\cite{googleAdMobAndroidAppOpen,googleAdMobIosAppOpen}.
\begin{table}[t]
\centering
\caption{Visual salience of trigger cues in touch- and motion-triggered splash advertisements.}
\label{tab:visual_salience}
\small
\renewcommand{\arraystretch}{1.2}
\setlength{\tabcolsep}{5pt}
\begin{tabular}{lcc}
\toprule
\textbf{Visual Dimension} &
\textbf{Touch-Triggered} &
\textbf{Motion-Triggered} \\
\midrule
Contrast            & High   & Low   \\
Text Size           & Large  & Small \\
Spatial Separation  & High   & Low   \\
Visual Guidance     & Strong & Weak  \\
\bottomrule
\end{tabular}
\end{table}

Background-app management may also contribute by affecting whether reopening an app resumes an existing page or starts a new session. Its effect depends on the app's advertising logic. Returning to an existing session can still trigger another ad, while a fresh launch may display none. App-open advertising supports foreground presentation on both Android and iOS~\cite{googleAdMobAndroidAppOpen,googleAdMobIosAppOpen}. Differences in display policies, advertising availability, and startup behavior are therefore plausible explanations for the observed patterns, although their individual contributions were not measured in our comparison.

\begingroup
\setlength{\emergencystretch}{3em}
\section{Existing Countermeasures}
\label{sec:defenses}

This section examines ad defense deployment from the user's perspective, including the software, permissions, and additional operations required. We compiled 108 countermeasure instances: 93 related to splash ads and 15 targeting feed, video, audio, and other ad formats. We classify them by where their defensive operations intervene and describe their scope and deployment requirements. An instance is a tool, rule set, version, prototype, or control, rather than necessarily an independent product. The inclusion criteria and counting rules are given in Appendix~\ref{app:defense_inventory}.

Another nine advertising SDK controls, two developer integration libraries, and one advertising-platform confirmation mechanism require action by host-app developers or advertising platforms. They are listed in Appendix~\ref{app:defense_cooperative}.

\textbf{Classification Basis.} We classify defensive operations into six categories by their intervention point. A tool can cover multiple stages: Fuck AD, for example, includes ad-loading interception, initialization prevention, and interface hiding or automatic skipping~\cite{fuckAd}. Of the 108 instances, 22 include operations spanning multiple stages and are counted in each corresponding category. Runtime injection, APK modification, and system configuration are implementation techniques. Ads may be preloaded, and motion monitoring may start before display, so the intervention stages need not follow a single strict sequence.

\textbf{Deployment Requirements.} Root or jailbreak denotes a system privilege condition, runtime injection modifies an app's behavior through a framework while it runs, and APK modification changes the installation package and requires reinstallation. We label these requirements separately. Xposed and LSPosed are runtime injection frameworks, which are not equivalent to APK modification. Accessibility, VPN, debugging, and overlay authorization provide different capabilities for cross-app control. Table~\ref{tab:defense_taxonomy} marks requirements for specific deployment paths and separately describes system settings, rules, certificates, source builds, and models. The appendix provides individual requirements and complete distributions.

\begin{table*}[!tp]
\centering
\caption{Deployment Paths, Requirements, and Estimated Effort for Advertising Defenses.}
\label{tab:defense_taxonomy}
\begingroup
\fontsize{7}{8.1}\selectfont
\setlength{\tabcolsep}{1.7pt}
\renewcommand{\arraystretch}{1.03}
\setlength{\aboverulesep}{.2ex}\setlength{\belowrulesep}{.2ex}
\renewcommand{\tabularxcolumn}[1]{m{#1}}
\begin{tabularx}{\textwidth}{@{}>{\raggedright\arraybackslash}m{2.5cm}>{\centering\arraybackslash}m{1.23cm}>{\centering\arraybackslash}m{1.04cm}>{\centering\arraybackslash}m{1.0cm}>{\centering\arraybackslash}m{.6cm}>{\centering\arraybackslash}m{1.15cm}>{\centering\arraybackslash}m{1.04cm}>{\centering\arraybackslash}m{.55cm}>{\centering\arraybackslash}m{.92cm}>{\raggedright\arraybackslash}m{2.9cm}>{\raggedright\arraybackslash}X@{}}
\toprule
\textbf{Deployment Path} & \textbf{Effort}
& \shortstack{\csname deficonroot\endcsname\\Root\\Jailbreak}
& \shortstack{\csname deficoninjection\endcsname\\Runtime\\Injection}
& \shortstack{\csname deficonapk\endcsname\\APK\\Mod.}
& \shortstack{\csname deficonaccessibility\endcsname\\Accessi-\\bility}
& \shortstack{\csname deficondebug\endcsname\\Debug-\\ging}
& \shortstack{\csname deficonvpn\endcsname\\VPN}
& \shortstack{\csname deficonoverlay\endcsname\\Overlay}
& \textbf{Other Permissions and Setup}
& \textbf{Representative Approaches and Fees} \\
\midrule[.65pt]
\multicolumn{11}{@{}l@{}}{\textbf{Blocking and Modifying Ad Resources (58)}} \\
\cmidrule[.35pt]{1-11}
System DNS configuration & \deploylow &  &  &  &  &  &  &  & \defreq{settings}{System Setting}; DNS address or configuration profile & AdGuard DNS (partly paid)~\cite{adGuardDnsService} \\
\cmidrule[.18pt]{1-11}
Local VPN domain filtering & \deploylow &  &  &  &  &  & $\checkmark$ &  & Built-in advertising-domain lists & AdAway (VPN)~\cite{adaway} \\
\cmidrule[.18pt]{1-11}
Root hosts filtering & \deployhigh & $\checkmark$ &  &  &  &  &  &  & Advertising-domain lists & AdAway (root)~\cite{adaway} \\
\cmidrule[.18pt]{1-11}
iOS HTTPS response rewriting & \deploymoderate &  &  &  &  &  & $\checkmark$ &  & Proxy, rules, certificate trust & Surge + app2smile (paid client)~\cite{app2smileSurgeAdsense,surgeIosLicenseFees} \\
\cmidrule[.18pt]{1-11}
In-app ad data / cache processing & \deployhigh & $\checkmark$ & $\checkmark$ &  &  &  &  &  & Framework, module, scope & MiFitnessAdAway~\cite{miFitnessAdAwaySource1} \\
\cmidrule[.18pt]{1-11}
Modifying APK loading logic & \deployhigh &  &  & $\checkmark$ &  &  &  &  & Apply patches, re-sign, install & Adobo patches~\cite{adoboAdsPatches} \\
\midrule[.65pt]
\multicolumn{11}{@{}l@{}}{\textbf{Preventing Ad Initialization and Startup (22)}} \\
\cmidrule[.35pt]{1-11}
Android entry-point interception & \deployhigh & $\checkmark$ & $\checkmark$ &  &  &  &  &  & Framework, module, scope & Fuck AD~\cite{fuckAd} \\
\cmidrule[.18pt]{1-11}
iOS entry-point interception & \deployhigh & $\checkmark$ & $\checkmark$ &  &  &  &  &  & Compatible tweak-loading environment & Chaoxing\allowbreak LaunchAdBlock~\cite{chaoxingLaunchAdBlock,chaoxingTweakDeployment} \\
\cmidrule[.18pt]{1-11}
Applying APK patches & \deployhigh &  &  & $\checkmark$ &  &  &  &  & Patch selection, re-signing, installation & Adobo advertising patches~\cite{adoboAdsPatches} \\
\cmidrule[.18pt]{1-11}
Installing a premodified APK & \deploymoderate &  &  & $\checkmark$ &  &  &  &  & Match version / signature; replace target app & CAD Viewer patch~\cite{cadViewerAdCleanerSource2} \\
\midrule[.65pt]
\multicolumn{11}{@{}l@{}}{\textbf{Suppressing Ad Presentation and Automatically Dismissing Ads (44)}} \\
\cmidrule[.35pt]{1-11}
Auto-dismissal with built-in rules & \deploylow &  &  &  & $\checkmark$ &  &  &  & Background-execution settings & Li Tiaotiao~\cite{liTiaotiao} \\
\cmidrule[.18pt]{1-11}
Auto-dismissal with imported rules & \deploymoderate &  &  &  & $\checkmark$ &  &  &  & Import subscription; background use & GKD + Lin-arm rules~\cite{gkdAdRules,linArmGkdRules} \\
\cmidrule[.18pt]{1-11}
Screenshot-based dismissal & \deploylow &  &  &  & $\checkmark$ &  &  &  & Screen access; built-in recognition; background use & obaby; Ad Skipper (no added VLM)~\cite{skipAdsObabySource1,madeyeAdSkipper} \\
\cmidrule[.18pt]{1-11}
Injection-based interface hiding & \deployhigh & $\checkmark$ & $\checkmark$ &  &  &  &  &  & Framework, module, scope & MinMinGuard; Fuck AD~\cite{minMinGuard,fuckAd} \\
\cmidrule[.18pt]{1-11}
Audio-ad muting & \deploylow &  &  &  &  &  &  &  & \defreq{notification}{Notification Access}; enable audio-ad muting & ad-free~\cite{adFreeAudio,adFreeAudioDelivery} \\
\midrule[.65pt]
\multicolumn{11}{@{}l@{}}{\textbf{Restricting Inputs and Disabling Triggers (9)}} \\
\cmidrule[.35pt]{1-11}
Vendor motion permissions & \deploylow &  &  &  &  &  &  &  & \defreq{settings}{System Setting}; supported device / firmware & Huawei; OPPO~\cite{huaweiShakeAdHelp,oppoMotionPermissionGuide} \\
\cmidrule[.18pt]{1-11}
Regular overlay touch interception & \deploymoderate &  &  &  &  &  &  & $\checkmark$ & User-selected region & One Patch (overlay path)~\cite{onePatchTouch} \\
\cmidrule[.18pt]{1-11}
Accessibility-overlay touch interception & \deploymoderate &  &  &  & $\checkmark$ &  &  &  & Select a region; alternative overlay path & One Patch (accessibility path)~\cite{onePatchTouch} \\
\cmidrule[.18pt]{1-11}
Temporary sensor control via debugging & \deploymoderate &  &  &  & $\checkmark$ & $\checkmark$ &  &  & Shizuku pairing; background use & NoShakingAD~\cite{noShakingAd} \\
\cmidrule[.18pt]{1-11}
Sensor listener interception & \deployhigh & $\checkmark$ & $\checkmark$ &  &  &  &  &  & Framework, module, scope & Fuck Shake; AdClose 4.3.7~\cite{fuckShake,adClose} \\
\midrule[.65pt]
\multicolumn{11}{@{}l@{}}{\textbf{Intercepting Ad Navigation and Requiring Confirmation (3)}} \\
\cmidrule[.35pt]{1-11}
User-enabled navigation reminder & \deploylow &  &  &  &  &  &  &  & \defreq{settings}{System Setting}; supported device & Honor navigation reminder~\cite{honorAdNavigationReminder} \\
\cmidrule[.18pt]{1-11}
Injection-based link / component interception & \deployhigh & $\checkmark$ & $\checkmark$ &  &  &  &  &  & Framework, module, matching rules & FuckAdJump; Li Tiansuo~\cite{fuckAdJumpSource,liTianSuo} \\
\midrule[.65pt]
\multicolumn{11}{@{}l@{}}{\textbf{Automatically Returning After Navigation (1)}} \\
\cmidrule[.35pt]{1-11}
Automatic return after navigation & \deployhigh &  &  &  & $\checkmark$ &  &  &  & \defreq{build}{Source Build}; rules; background use & ShakeGuard (prototype)~\cite{shakeGuardSource,shakeGuardSourceDelivery} \\
\bottomrule
\end{tabularx}
\par\smallskip
\begin{minipage}{\textwidth}
\fontsize{6.7}{8}\selectfont
\raggedright
\textbf{Deployment Effort.}
Deployment effort is rated qualitatively based on the technical skills and setup steps required of ordinary users.
\deploylow{} installation and authorization, or an existing system setting;
\deploymoderate{} rules, region selection, certificates, debugging, or replacement-app setup;
\deployhigh{} root / jailbreak, runtime instrumentation, APK patching, or source compilation.

\par\smallskip
\textbf{Deployment Symbols.}
\defreq{root}{Root / Jailbreak};
\defreq{injection}{Runtime Injection};
\defreq{apk}{APK Modification};
\defreq{accessibility}{Accessibility};
\defreq{debug}{Debugging};
\defreq{vpn}{VPN};
\defreq{overlay}{Overlay};
\defreq{settings}{System Setting};
\defreq{build}{Source Build};
\defreq{model}{Additional Model};
\defreq{notification}{Notification Access}.
The key denotes root on Android and jailbreaking on iOS.
\par\smallskip
\textbf{Requirement Indicators.}
A checkmark denotes a requirement of the listed path; blanks do not mean no setup is needed. Other permissions appear in their own column. The APK column includes premodified apps. Alternative paths are separate rows.
\par\smallskip
\textbf{Counts and Fees.}
Category counts cover 108 instances; 22 span multiple stages. Splitting paths does not add instances. As of September 13, 2026, ``paid'' requires a software / service purchase; ``partly paid'' combines a free path with paid features. Details appear in Appendix~\ref{app:defense_distribution}.
\end{minipage}
\endgroup
\end{table*}

\subsection{Ad Defense Methods}
\label{sec:defense_methods}

We describe defensive operations, user deployment requirements, and principal limitations across the six intervention stages.

\subsubsection{Blocking and Modifying Ad Resources}

These methods prevent apps from obtaining usable ad content by blocking requests, modifying responses and configurations, or restricting local resource reads. Users can choose a \defreq{vpn}{VPN} client or \defreq{settings}{System Setting} for DNS or network management. Modifying HTTPS content additionally requires a proxy and certificate trust, while in-app processing may require \defreq{injection}{Runtime Injection} or \defreq{apk}{APK Modification}. Of the 58 instances in this category, 36 use network tools, DNS, or vendor network-management configurations, and 22 involve app internals or system files.

\textbf{Blocking Network Ad Requests.} Domain filtering refuses to resolve ad server addresses, while URL filtering additionally examines request paths. For example, AdAway filters using advertising-domain rules, and R-Store provides rules for blocking specific ad requests~\cite{adaway,rStoreMstandConfiguration}. When blocking succeeds, users do not need separate protection configurations for taps, swipes, or shakes. However, a domain may also serve normal data, and rules require updates as server interfaces change.

Filtering can run on the device or at a remote resolver. DNS66, personalDNSfilter, and Blokada 5 provide local paths, while Blokada Cloud and the inspected Block This implementation use remote filtering services. The former require a configured local executor, while the latter depend on the corresponding service~\cite{dnsSixtySix,personalDnsFilter,blokadaWebsite,blockThis}. TrackerControl processes connections through a \defreq{vpn}{VPN} and blocks requests according to domain associations, advertising categories, and user choices~\cite{trackerControlAdsSource}. On supported iOS and iPadOS versions, users can enable Wipr 2's Filtr system URL filtering extension. Coverage still depends on requests passing through supported paths~\cite{wiprFiltr,appleNetworkExtensionURLFilters}. Classifiers can also identify ad requests in place of individual rules: NoMoAds, for example, uses network and visible-content features to classify advertising traffic~\cite{shuba2018nomoads}. It is retained in Appendix~\ref{app:defense_limited} outside the main-inventory counts because its installation or complete build route is unresolved.

\textbf{Modifying Responses and Configurations.} These methods allow requests to complete, then remove advertising entries from returned data or modify ad status. Clients such as Surge and Quantumult X execute app2smile's ad-network rules to modify status or result fields at specified endpoints. blackmatrix7's rules also modify display duration, dimensions, and validity periods~\cite{app2smileSurgeAdsense,app2smileQuantumultAdsense,blackmatrixStartup}. Users need a compatible client and rules. The listed iOS proxy path additionally requires a \defreq{vpn}{VPN}, and HTTPS content modification requires the app to accept the proxy certificate. Apps that trust only prespecified certificates may reject the connection~\cite{surgeHttpsMitm}. Returned data can also be processed inside an app: BetterHeybox uses \defreq{injection}{Runtime Injection} to remove advertising entries from content lists. This path requires loading a module but does not require decrypting responses in a network proxy~\cite{betterHeyboxSource1}. The same method applies to feed and recommendation ads. Discover Feed Filter filters advertising data through injection, while BiliRoamingX handles recommendation ads through APK patches. The latter's referenced prebuilt distribution channel is unavailable, and its patch-building and installation channels remain unconfirmed~\cite{discoverAdsFilter,biliRoamingXAds,biliRoamingXAdsDelivery}. BiliRoamingX is therefore retained in Appendix~\ref{app:defense_limited} outside the main-inventory counts.

\textbf{Restricting Local Ad Resource Reads.} Filtering new requests does not prevent an app from reading cached ads or ads bundled in its installation package. MiFitnessAdAway replaces the result of reading the splash-ad cache inside the app with an empty value. Its listed deployment path requires \defreq{root}{Root} and \defreq{injection}{Runtime Injection}~\cite{miFitnessAdAwaySource1}. Its protection is limited to matching cache-read paths.

These methods must distinguish advertising from normal data and cover the resource sources actually used. Unmatched rules, readable caches, or a switch in ad sources can allow ads to continue appearing.

\subsubsection{Preventing Ad Initialization and Startup}

These methods disable ad initialization or startup functions, or bypass or block ad-page launches. Users typically need a \defreq{injection}{Runtime Injection} framework and an advertising module, or a target app installed through \defreq{apk}{APK Modification}. A root-based framework also requires \defreq{root}{Root}. All 22 instances in this category involve injection or APK modification: 18 provide injection paths, five provide APK modification paths, and one provides both. Two require source builds.

\textbf{Disabling Ad Initialization or Startup Functions.} Runtime injection can replace a specified function's behavior when an app calls it, making the function return immediately. For example, FanqieHook intercepts splash entry points in supported apps, and Fuck AD intercepts initialization of supported advertising SDKs~\cite{fanqieHookSource1,fuckAd}. An SDK is an advertising library integrated into the host app, and users intervene in its execution by installing a module. They need to install a compatible framework, enable the module, and select the target apps in its scope.

On iOS, ChaoxingLaunchAdBlock also intercepts advertising startup code in Chaoxing Xuexitong. User deployment requires jailbreaking and a corresponding tweak-loading environment~\cite{chaoxingLaunchAdBlock,chaoxingTweakDeployment}. Another path modifies the installation package: Adobo's advertising patches make specified SDK display entry points return early. Users can directly use compiled patches, modify an APK through a manager, and install it~\cite{adoboAdsPatches}.

\textbf{Bypassing or Blocking Ad-Page Launches.} The CAD Viewer patch changes the launch entry point to the home page, and users can install the modified APK provided by the project. CoolApkNoSplash instead intercepts launch requests targeting CoolApk's splash page at runtime and requires an injection framework and the corresponding module~\cite{cadViewerAdCleanerSource2,coolApkNoSplash}. The former installs a modified app; the latter loads a module while the app runs.

These operations depend on accurately identifying ad entry points. Existing adaptations may fail when functions, call patterns, or startup order change. If advertising and normal initialization share functions, skipping an entire function may also affect host-app functionality.

\subsubsection{Suppressing Ad Presentation and Automatically Dismissing Ads}

These methods reduce advertising interruptions by hiding ad elements, operating close controls, or muting audio. Automatic dismissal typically requires \defreq{accessibility}{Accessibility}, and visual recognition also involves screen access and local computation. Interface hiding typically uses \defreq{injection}{Runtime Injection} or \defreq{apk}{APK Modification}, while audio-ad muting uses notification access and audio control. Of the 44 instances in this category, 24 provide accessibility paths and 20 provide injection or package-modification paths, with one providing both. Another uses notification access and audio control. Seven require source builds, and one has paid features.

\textbf{Hiding or Removing Ad Elements.} MinMinGuard sets ad View heights to zero, while MiFitnessAdAway hides the corresponding ad interface and removes its occupied space~\cite{minMinGuard,miFitnessAdAway}. The listed Android injection paths require \defreq{root}{Root} and \defreq{injection}{Runtime Injection}. Users must enable the module and match the target version. Fuck AD provides both injection-based interface hiding and accessibility-based automatic skipping, requiring an injection framework and accessibility authorization, respectively~\cite{fuckAd}. Some automatic-click paths additionally use Shizuku and require \defreq{debug}{Debugging}. The same interface-hiding mechanism is used for other ad formats: GmailHideAds hides advertising controls in email lists, and amznkiller injects styles into Amazon Shopping's web container to hide sponsored content. Both require a module loaded in the target app~\cite{gmailHideAds,amznkillerSource}.

\textbf{Automatically Dismissing Ads.} Once authorized, an Android accessibility service can read interface information exposed by other apps and perform clicks. Li Tiaotiao, TapClick, GKD, and its advertising subscriptions locate skip buttons using text, positions, or control identifiers and operate them for the user. GKD's specific coverage is determined by its subscription rules~\cite{androidAccessibilityService,liTiaotiao,tapClick,gkdAdRules,linArmGkdRules}. Users need \defreq{accessibility}{Accessibility}, rule configuration as required by the tool, and background execution. They typically do not need to modify the target app. \textbf{\textit{Android's accessibility-based auto-skipping does not have a directly equivalent deployment path for ordinary third-party tools on iOS.} App isolation and code signing restrict those tools from reading and operating other apps in the same way~\cite{appleRuntimeSecurity,appleCodeSigningSecurity}}.

When an ad does not expose an independently operable control, its close button can be recognized from a screenshot. obaby's tool captures screenshots, uses object detection to locate buttons, and clicks through an accessibility service. Ad Skipper combines interface-node, text, and image recognition~\cite{skipAdsObabySource1,madeyeAdSkipper}. Both require screen access and recognition computation, and Ad Skipper also offers an optional configuration with an additional downloaded VLM. User-configured visual rules also support video-ad dismissal: Klick'r matches image and text conditions, and users have reported using it to recognize and automatically click close buttons after game video ads end~\cite{klickrOfficialFeatures,klickrAdCloseReport}.

\textbf{Muting Audio Ads.} ad-free identifies audio ads from playback states in supported apps and reduces interruptions by lowering the volume or playing replacement audio. Users need to grant \defreq{notification}{Notification Access} and allow the corresponding audio control. They can directly install the package provided by F-Droid~\cite{adFreeAudio,adFreeAudioDelivery}.

Automatic dismissal must first detect the ad and act, before which the ad may still trigger. Interface changes or deceptive close buttons may also cause clicks to land in the ad region. Hiding the interface may leave background advertising logic running, and muting does not shorten the audio ad's playback time. Accessibility authorization also gives the tool access to non-ad interfaces.

\subsubsection{Restricting Inputs and Disabling Triggers}

These methods restrict ads' access to touch or motion input through touch-region overlays, vendor motion permissions, or sensor restrictions implemented through debugging services or injection modules. Of the nine instances in this category, four are user-enabled system settings, three use injection, one controls sensors through Shizuku debugging authorization, and one provides an overlay. Overlays can use overlay or accessibility paths, each requiring its corresponding authorization.

\textbf{Intercepting Touch Input.} One Patch lets users select a rectangular region and uses an overlay to receive touches within it, preventing events from reaching the ad. Its regular overlay path requires \defreq{overlay}{Overlay}, while its accessibility-overlay path requires \defreq{accessibility}{Accessibility}~\cite{onePatchTouch}. Users must select the region themselves. Normal controls inside it are also blocked, while touches outside it and motion input remain unaffected.

\textbf{Restricting Motion Sensor Access.} On supported devices, Huawei, OPPO, and Honor users can deny device-orientation or motion permissions to a target app through \defreq{settings}{System Setting}. Some vivo firmware also provides an option to deny access only during splash ads~\cite{huaweiShakeAdHelp,oppoMotionPermissionGuide,honorMotionPermissionRelease,vivoOriginOSLaunchMotionRelease}. These are user-enabled system controls whose protection depends on the device model, firmware, and permission implementation.

Third-party tools provide two different deployment paths. NoShakingAD requires \defreq{accessibility}{Accessibility} and \defreq{debug}{Debugging}: accessibility events observe the foreground app, while Shizuku provides debugging authorization to switch the global sensor mode and schedule restoration after five seconds~\cite{noShakingAd,noShakingServiceRecovery}. Fuck Shake, the verified AdClose 4.3.7 configuration, and QzxyAdBlock restrict sensor listeners through \defreq{root}{Root} and \defreq{injection}{Runtime Injection}~\cite{fuckShake,adClose,qzxyAdBlock}. Fuck Shake intercepts one listener-registration interface. Other interfaces and previously established listeners require separate handling.

Sensor restrictions also affect normal motion-based functions, and global restrictions involve other apps. Ads still displayed after a temporary protection period may regain access to motion input. Input control does not directly dismiss ads, and restricting motion does not automatically prevent tap- or swipe-triggered navigation.

\subsubsection{Intercepting Ad Navigation and Requiring Confirmation}

These methods rewrite destination links, deny destination-page launches, or require user confirmation when an ad is about to open a web page or another app. Of the three instances in this category, two use \defreq{root}{Root} and \defreq{injection}{Runtime Injection}, and one uses \defreq{settings}{System Setting} on supported devices.

\textbf{Rewriting Links and Intercepting Page Launches.} When an app parses a URI, FuckAdJump replaces strings matching Taobao or JD scheme prefixes with an empty string. Li Tiansuo's UC rules intercept launches of ad landing pages or browser Activities specified in the rule list~\cite{fuckAdJumpSource,liTianSuo}. Users need a compatible injection framework, a module, and its target-app scope. Protection is determined by the matching links and components.

\textbf{Requiring User Confirmation for Navigation.} Honor's automatic app redirection reminder lets users enable a setting on supported systems to display a reminder for recognized navigation caused by accidental ad interactions. Official documentation includes third-party app launches caused by shaking, swiping, and tapping~\cite{honorAdNavigationReminder,honorEsg2024AdNavigation}. Users can enable the setting directly without installing an injection framework.

Navigation control requires requests to pass through the configured checkpoint and may also affect shopping, login, or payment links that users intentionally open. Ads can continue to display. Denying an external launch does not establish that other paths, such as in-app web pages, have also been blocked.

\subsubsection{Automatically Returning After Navigation}

These methods detect a transition after the destination app has opened and attempt to return, shortening the interruption. The one included instance is ShakeGuard, which requires a \defreq{build}{Source Build}, \defreq{accessibility}{Accessibility}, source and destination rules, and background execution.

ShakeGuard combines accessibility window events, app-matching rules, and a protection period to determine whether a transition needs handling, then sends a system Back operation. If returning fails, it can attempt to reopen the source app~\cite{shakeGuardSource}. Users need JDK 17, Android SDK 35, and Gradle, then build and install the APK before granting authorization and configuring rules~\cite{shakeGuardSourceDelivery}. The defense tool itself is compiled; the host APK remains unchanged.

Recovery occurs after navigation and cannot undo pages already displayed or network requests already sent by the destination app. It also does not guarantee a return to the original host-app page. Source, destination, and time-window information provide only indirect clues, so normal transitions may also match the rules.

\subsection{Case Study: Empirical Evaluation of Representative Defenses}
\label{sec:defense_empirical}

We selected 13 configurations of 11 tools and tested the ten host apps listed in Section 3 on a Pixel 6 and a Xiaomi MIX 2S. Each cell in Table~\ref{tab:defense_empirical} represents a configuration and app combination, summarizing results from both devices. Deployment labels in the table describe the configurations actually tested. Vector is the runtime code-injection framework used in these tests, and Shizuku provides debugging authorization for the corresponding configuration.

\begin{table*}[!tp]
\centering
\caption{Empirical Results for 13 Configurations of 11 Defense Tools Across 10 Tested Apps.}
\label{tab:defense_empirical}
\begingroup
\fontsize{7}{8.3}\selectfont
\setlength{\tabcolsep}{2pt}
\renewcommand{\arraystretch}{1.15}
\renewcommand{\tabularxcolumn}[1]{m{#1}}
\begin{tabularx}{\textwidth}{@{}>{\raggedright\arraybackslash}m{3.5cm}>{\centering\arraybackslash}m{1.8cm}*{10}{>{\centering\arraybackslash}X}@{}}
\toprule
\textbf{Tool / Configuration} & \shortstack{\textbf{Deployment}\\\textbf{Requirements}} & \shortstack{\textbf{Sogou}\\\textbf{Input}} & \shortstack{\textbf{Baidu}} & \shortstack{\textbf{Baidu}\\\textbf{Maps}} & \shortstack{\textbf{Baidu}\\\textbf{Input}} & \shortstack{\textbf{Weibo}} & \shortstack{\textbf{QQ}\\\textbf{Browser}} & \shortstack{\textbf{Tencent}\\\textbf{Video}} & \shortstack{\textbf{iQIYI}} & \shortstack{\textbf{Tencent}\\\textbf{News}} & \shortstack{\textbf{Mango}\\\textbf{TV}} \\
\midrule
AdAway (VPN)~\cite{adaway} & \csname deficonvpn\endcsname & $\times$ & $\checkmark$ & $\times$ & $\checkmark$ & $\times$ & $\times$ & $\times$ & $\times$ & $\times$ & $\checkmark$ \\
\cmidrule[.18pt]{1-12}
AdAway (root) & \csname deficonroot\endcsname & $\times$ & $\triangle$ & $\triangle$ & $\triangle$ & $\times$ & $\times$ & $\times$ & $\times$ & $\times$ & $\times$ \\
\cmidrule[.18pt]{1-12}
TrackerControl (ad blocking)~\cite{trackerControlAndroidDocs} & \csname deficonvpn\endcsname & $\checkmark$ & $\checkmark$ & $\checkmark$ & $\checkmark$ & $\times$ & $\times$ & $\times$ & $\times$ & $\times$ & $\checkmark$ \\
\cmidrule[.18pt]{1-12}
AdClose~\cite{adClose} & \csname deficonroot\endcsname\quad \csname deficoninjection\endcsname & $\times$ & $\diamond$ & $\triangle$ & $\triangle$ & $\times$ & $\diamond$ & $\diamond$ & $\times$ & $\times$ & $\times$ \\
\cmidrule[.18pt]{1-12}
Fuck AD~\cite{fuckAd} & \csname deficonroot\endcsname\quad \csname deficoninjection\endcsname & $\times$ & $\triangle$ & $\triangle$ & $\triangle$ & $\times$ & $\checkmark$ & $\times$ & $\checkmark$ & $\times$ & $\times$ \\
\cmidrule[.18pt]{1-12}
Li Tiaotiao 2.2~\cite{liTiaotiao} & \csname deficonaccessibility\endcsname & $\checkmark$ & $\checkmark$ & $\checkmark$ & $\checkmark$ & $\checkmark$ & $\checkmark$ & $\checkmark$ & $\checkmark$ & $\checkmark$ & $\checkmark$ \\
\cmidrule[.18pt]{1-12}
GKD + Lin-arm rules~\cite{gkdAdRules,linArmGkdRules} & \csname deficonaccessibility\endcsname & $\times$ & $\times$ & $\times$ & $\times$ & $\times$ & $\times$ & $\times$ & $\times$ & $\times$ & $\times$ \\
\cmidrule[.18pt]{1-12}
Ad Skipper (without VLM)~\cite{madeyeAdSkipper} & \csname deficonaccessibility\endcsname & $\checkmark$ & $\triangle$ & $\triangle$ & $\triangle$ & $\times$ & $\checkmark$ & $\times$ & $\checkmark$ & $\checkmark$ & $\checkmark$ \\
\cmidrule[.18pt]{1-12}
Ad Skipper (with VLM) & \csname deficonaccessibility\endcsname\quad \csname deficonmodel\endcsname & \textemdash{} & \textemdash{} & \textemdash{} & \textemdash{} & \textemdash{} & \textemdash{} & \textemdash{} & \textemdash{} & \textemdash{} & \textemdash{} \\
\cmidrule[.18pt]{1-12}
NoShakingAD~\cite{noShakingAd} & \csname deficonaccessibility\endcsname\quad \csname deficondebug\endcsname & $\times$ & $\times$ & $\times$ & $\times$ & $\times$ & $\times$ & $\checkmark$ & $\checkmark$ & $\times$ & $\times$ \\
\cmidrule[.18pt]{1-12}
Fuck Shake~\cite{fuckShake} & \csname deficonroot\endcsname\quad \csname deficoninjection\endcsname & $\times$ & $\triangle$ & $\triangle$ & $\triangle$ & $\times$ & $\checkmark$ & $\times$ & $\times$ & $\times$ & $\times$ \\
\cmidrule[.18pt]{1-12}
FuckAdJump~\cite{fuckAdJumpSource} & \csname deficonroot\endcsname\quad \csname deficoninjection\endcsname & $\times$ & $\triangle$ & $\triangle$ & $\triangle$ & $\times$ & $\times$ & $\times$ & $\times$ & $\times$ & $\times$ \\
\cmidrule[.18pt]{1-12}
ShakeGuard~\cite{shakeGuardSource} & \csname deficonbuild\endcsname\quad \csname deficonaccessibility\endcsname & \textemdash{} & \textemdash{} & \textemdash{} & \textemdash{} & \textemdash{} & \textemdash{} & \textemdash{} & \textemdash{} & \textemdash{} & \textemdash{} \\
\bottomrule
\end{tabularx}
\par\smallskip
\begin{minipage}{\textwidth}
\fontsize{6.7}{8}\selectfont
\raggedright
$\checkmark$: prevented navigation triggered by the target ad; $\times$: failed to prevent it; $\triangle$: not evaluated because the target splash ad was not observed with the defense disabled; $\diamond$: the host app failed to launch or became stuck with the defense enabled; \textemdash{}: not successfully validated.

The tested injection framework was Vector 2.2, which requires root. Shizuku provides debugging authorization for NoShakingAD. The additional-VLM configuration did not run successfully; the configuration without an additional VLM still performs built-in local inference. ShakeGuard requires compiling the defense tool itself.

\textbf{Deployment Icons.} \csname deficonvpn\endcsname VPN; \csname deficonroot\endcsname Root; \csname deficoninjection\endcsname Runtime Injection; \csname deficonaccessibility\endcsname Accessibility; \csname deficonmodel\endcsname Additional Model; \csname deficondebug\endcsname Debugging; \csname deficonbuild\endcsname Source Build.

\textbf{Configuration Details.} AdAway (root): Modifies system hosts; Li Tiaotiao 2.2: Background execution; GKD + Lin-arm rules: Lin-arm rules, background execution; Ad Skipper (without VLM): Screen access, built-in inference; Ad Skipper (with VLM): Additional VLM + visual projection files; NoShakingAD: Shizuku, background execution; ShakeGuard: Build toolchain, source and destination rules.
\end{minipage}
\endgroup
\end{table*}

\textbf{Successful Protection Covered Only a Minority of Combinations.} Of the 130 combinations, 28 (21.5\%) succeeded, 62 failed, 17 had no observed baseline ad, three experienced host-app failures, and 20 belonged to configurations that were not successfully validated. Success means preventing navigation triggered by the target ad.

\textbf{Ad Delivery Limited Evaluation Coverage.} Baidu, Baidu Maps, and Baidu Input displayed no splash ads on the Pixel 6 even with defenses disabled, suggesting that delivery may depend on conditions such as the device environment. We did not count this absence of ads as successful protection.

\textbf{Network Filtering Left Multiple Apps Unprotected.} AdAway's \defreq{vpn}{VPN} mode succeeded in three of ten apps, and TrackerControl's \defreq{vpn}{VPN} configuration succeeded in five. AdAway's \defreq{root}{Root} mode succeeded in none of the seven evaluable apps. Domain filtering depends on rules, while app-open ad SDKs support preloading. Uncovered domains and ads cached before filtering was enabled are possible failure paths~\cite{adaway,googleAdMobAndroidAppOpen}.

\textbf{Accessibility Tools' Effectiveness Varied by Configuration.} Li Tiaotiao 2.2, GKD with Lin-arm rules, and Ad Skipper without an additional VLM all use \defreq{accessibility}{Accessibility}. They succeeded in all ten apps, none of the ten apps, and five of seven evaluable apps, respectively. Missing rules or interface changes may prevent target location~\cite{gkdAdRules,linArmGkdRules}. In manual observations, ads remained visible for approximately one second before Li Tiaotiao dismissed them. Ad Skipper's node matching, OCR, and built-in YOLO also require corresponding local computation.

\textbf{Root and Injection Do Not Guarantee Effectiveness or Compatibility.} The tested configurations of Fuck AD, Fuck Shake, FuckAdJump, and AdClose all require \defreq{root}{Root} and \defreq{injection}{Runtime Injection}, together producing only three successful combinations. With AdClose enabled, Baidu could not launch, while QQ Browser and Tencent Video remained on the ad page. Its 4.3.7 configuration enabled sensor restrictions and request inspection, had no network-blocking rules, and did not implement the SDK initialization interception described in the old README.

\textbf{Motion Protection Leaves Touch Triggers and Ad Display Intact.} NoShakingAD's \defreq{accessibility}{Accessibility} and \defreq{debug}{Debugging} configuration can prevent motion-triggered navigation but does not dismiss the ad or prevent touch-triggered navigation. According to the trigger categories listed in Section 3, Tencent Video and iQIYI have only motion triggers and are therefore marked successful in the table. Apps with touch triggers are marked unsuccessful even when their motion-triggered navigation was prevented.

\textbf{Model Initialization Failed Before Ad Recognition.}
On the Pixel 6 running Android 13, Ad Skipper v1.3's \defreq{accessibility}{Accessibility} and \defreq{model}{Additional Model} configuration crashed while loading either InternVL3 2B or Qwen2.5-VL 3B, despite complete downloads of the models and their visual projection files. Both failures occurred at the same Vulkan buffer-allocation call before ad recognition began. The configuration without an additional VLM could still run.

\textbf{Source-Level Tests Revealed State-Management Defects in ShakeGuard.}
ShakeGuard requires a \defreq{build}{Source Build} and \defreq{accessibility}{Accessibility}; its on-device deployment was not completed. We separately executed its original state-management code on the JVM with controlled foreground events and return outcomes, reproducing two defects. After a failed return, the prototype could remain in its recovery state and stop evaluating transitions from other protected apps. Returning to the launcher and reopening the same app could also retain the previous protection-window start time, causing a new redirect to be allowed as outside the window~\cite{shakeGuardSource}.

\endgroup

\section{Why So Challenging to Stop Splash Ads}
\label{sec:gaps}

Section~\ref{sec:defenses} shows that most existing countermeasure tools show unreliable performance, even with technical resource requirements for deployment that far exceed the capability of average smartphone users. Unfortunately, most of these tools' capabilities do not always translate into protection that users can continue to use. This section draws on existing mechanisms and our empirical results to analyze how these difficulties affect defenses.

\subsection{Difficulty Separating Advertising from Normal Functionality}

Advertising code is integrated into the host application and shares application capabilities and navigation paths with normal functionality. Defenses need to block advertising behavior while preserving host functionality, but the scope defined by application permissions, internal functions, and destination links does not always align with the scope of advertising behavior.

\textbf{Restricting capabilities per application also affects normal functionality.} Restricting motion-sensor access per application can prevent ads from obtaining shake data, but also limits motion-based interactions within the same application~\cite{huaweiShakeAdHelp}. This control covers the entire application and cannot deny sensor access only to advertising code.

\textbf{Selective interception depends on accurately identifying advertising code.} Disabling only ad initialization functions or restricting only sensor listeners registered by ads can reduce the impact on normal functionality. However, the defense must accurately locate these entry points and accommodate how they are invoked. For example, Fuck AD's interception of ad SDK initialization requires matching the corresponding functions in supported SDKs~\cite{fuckAd}. This control requires adaptation to individual advertising implementations.

\textbf{Interception based on destination links has difficulty distinguishing advertising from normal navigation.} FuckAdJump blocks external app launches based on Taobao and JD link prefixes, but the same prefixes can also be used in shopping links that users intentionally open~\cite{fuckAdJumpSource}. Removing ads that have already been identified does not require determining whether users intended to trigger them; the distinction needed is between advertising and normal operations that share the same navigation path. HONOR's navigation reminders request an additional confirmation, allowing users to decide whether to proceed~\cite{honorAdNavigationReminder}. This approach adds an interaction, and its coverage also depends on which navigation attempts enter the confirmation process.

\subsection{Intervention Timing and Execution Paths Limit Coverage}

A tool's ability to intervene at one stage of the advertising process does not mean that it can address every trigger type before navigation occurs. Protection depends both on when the defense takes effect and on whether the actual resource, input, and navigation paths pass through it.

\textbf{Ads may still trigger navigation before a defense takes effect.} Automatic skipping must first detect the ad, locate its dismissal control, and then perform a click. Although Li Tiaotiao succeeded in all ten applications in our tests, ads remained visible for approximately one second before dismissal. During this interval, the ad has not yet exited, and its touch or motion trigger logic may still execute. Returning after navigation can attempt recovery only after the interruption has occurred. These operations cannot continuously prevent navigation from the moment an ad appears.

\textbf{Advertising behavior that does not pass through the interception point can continue to execute.} Network filtering can block resources before an ad appears, but blocking new requests alone cannot address content that has already been cached; app-open advertising SDKs themselves also support preloading~\cite{googleAdMobAndroidAppOpen}. Input control can operate before a dismissal button appears, but our NoShakingAD results show that restricting motion triggers still leaves ad presentation and touch triggers intact. Navigation control also requires requests to pass through the intercepted interface and therefore cannot guarantee the same protection for other paths.

\subsection{Defenses Depend on Rule Coverage and Correct Configuration}

Defenses that rely on rules identify ads through predefined matching conditions. After a tool is installed, protection still requires rules to be correctly enabled, cover the target ad, and remain applicable to the particular ad served.

\textbf{Tools cannot handle the target ad when rules do not cover it or are not enabled.} Domain filtering must match advertising addresses, automatic skipping must match dismissal controls, and function interception must locate the corresponding advertising entry points. For example, GKD provides its execution capabilities separately from advertising rule subscriptions, which determine the applications and controls to match~\cite{gkdAdRules,linArmGkdRules}. Enabling the execution tool alone cannot compensate for ads that its rules do not cover. In the AdClose configuration we tested, request inspection was enabled, but no network-blocking rules were configured; inspecting requests does not itself block them.

\textbf{The same application version may exhibit different advertising behavior.} An example in AdHive shows an advertising SDK reading shake enablement and thresholds from a server response and using local defaults when the corresponding parameters are absent~\cite{wu2026adhive}. The installed application version therefore does not fully determine the ad's trigger conditions. Rule applicability also depends on the interface, configuration, and execution path of the particular ad presentation. Claiming support solely by application name cannot adequately express these differences in coverage; one successful attempt does not guarantee that subsequently served ads will meet the same conditions.

\subsection{Ordinary Users Face Deployment and Runtime Barriers}

Users need not only to obtain a tool but also to make the defense run correctly while keeping the host application usable. The additional privileges, build or model dependencies, and runtime failures associated with different approaches can all prevent users from obtaining effective protection.

\textbf{The privileges required by a defense may entail additional authorization or system modifications.} The approaches in Section~\ref{sec:defenses} variously require VPN or accessibility authorization, debugging privileges obtained through Shizuku, root access and runtime code-injection frameworks, or modification and reinstallation of the target application. Advertising code can execute within its host application, whereas an independent defense tool requires additional privileges to observe or change the behavior of other applications. These control capabilities often cannot be obtained simply by installing an application.

\textbf{Build and model environments can also prevent tools from running.} Tools that do not require root may still have additional dependencies. ShakeGuard requires users to configure a build environment and compile an installation package before enabling the accessibility service and rules~\cite{shakeGuardSource,shakeGuardSourceDelivery}. In our Pixel~6 tests, Ad Skipper crashed before ad recognition began when loading either of the two tested VLM models, even though the models and their visual projection files had been fully downloaded. Both failures occurred while the Vulkan graphics backend was allocating a buffer; the configuration without an additional VLM could still run.

\textbf{Problems in execution logic and compatibility may interrupt defenses or host functionality.} In tests executing ShakeGuard's original state-management code on the JVM, a failed return could leave the prototype in its recovery state, preventing it from evaluating subsequent navigation from other protected applications. Returning to the home screen and reopening the same application could also retain the previous timer start, incorrectly treating a new navigation attempt as outside the protection window~\cite{shakeGuardSource}. In our phone tests, enabling AdClose prevented Baidu from launching and left QQ Browser and Tencent Video stuck on the ad page.

\subsection{Fragmented Maintenance and Distribution Increase the Cost of Continued Use}

Continued use of advertising defenses requires users to obtain tools that are still maintained, select versions that include the required features, and keep rules compatible with target applications. Fragmented maintenance and distribution of tools and rules make these tasks difficult to complete through a single installation.

\textbf{Rule adaptation and project migration require users to keep track of updates.} Existing rules may need revision when advertising addresses, interfaces, or internal functions change, while tools, rules, and target applications may be released separately by different teams. Project migrations also change where updates are obtained: Ad-Cleaner has been archived and is no longer maintained, with subsequent functionality moving to R-Store~\cite{adCleanerMigration}. Users who continue using the old project need to find the subsequent maintenance source to obtain further compatibility updates.

\textbf{Diverse names and versions increase the burden of choosing an installation package.} A website introducing Li Tiaotiao downloads lists both ``Paidaxing'' 2.2 and 2.4 alongside similar tools such as Lei Tiaotiao and Yizhichan. The site identifies 2.2 as the original author's final release and describes 2.4 as a third-party modification. It also states that rules continue to be updated after software updates have ceased and provides instructions for manual import~\cite{leiTiaoTiao}. Users must distinguish versions of the same tool from separate projects and check maintenance sources, system compatibility, and accompanying rules. Names and version numbers alone make it difficult to determine which package to download; an unchanged software version also does not mean that its rules no longer require updates.

\textbf{Different distribution channels may also provide different advertising defense features.} NetGuard's project documentation explicitly states that its hosts-based ad filtering is not included in the Google Play release~\cite{netGuardAdsMode}. Thus, even after selecting a tool, users must verify that the installed version includes the features they need.

\subsection{Potential Legal Problems}
Overly intrusive ads drive users to seek ad-removal tools, but tools that indiscriminately remove all ads may affect applications' advertising-supported business models. Balancing reduced ad intrusion with the preservation of advertising revenue presents another challenge for existing defenses. Li Tiaotiao is free to use and was the only tool in our tests that prevented the target ad-triggered navigation across all ten applications. \textbf{\textit{However, its developer announced an indefinite suspension of updates in August 2023 after receiving a lawyer's letter.}} The letter alleged that the tool interfered with a browser's advertising business and constituted unfair competition. The developer emphasized that the tool merely clicked existing skip buttons on users' behalf and argued that difficult-to-use or misleading dismissal controls drove users to seek such tools~\cite{liTiaotiao2023SuspensionReport}.

\textbf{Protecting Users from Intrusion Does Not Necessarily Require Removing All Ads.} The Li Tiaotiao dispute illustrates that the parties may differ in their understanding of which behaviors a defense actually changes. Automatically clicking an existing skip button exercises an interface-provided exit option on the user's behalf. Restricting motion triggers or external-app navigation can preserve ad display while preventing unintended behavior. Blocking ad-resource loading directly affects opportunities to display ads. These approaches affect user experience and advertising businesses differently, but their impact cannot simply be ranked by technical mechanism: navigation interception that preserves ad display may still affect conversions, while automatic skipping may reduce the time users spend viewing ads. Thus, even defenses targeting intrusive interactions may not entirely avoid conflicts with advertising revenue.

\textbf{Restricting Defense Tools Alone Cannot Resolve This Tension.} If dismissal controls remain difficult to use and everyday movements can still trigger navigation, users retain an incentive to seek more comprehensive ad blocking. Conversely, indiscriminately removing all ads fails to accommodate services supported by nonintrusive advertising. Clearer, easier-to-use exit options and restrictions on unintended navigation may reduce users' demand for blanket ad blocking. However, making these protections effective while preserving sustainable advertising revenue still requires coordination among advertising providers, application developers, and defense-tool developers.

\section{Future Paths Toward Better Ecosystems}
\label{sec:future}
Finally, we provide some thoughts on the possible paths forward for better balance between ad providers' power and users' controls. 

\subsection{Technical Research Exploration for More Usable Third-Party Tools}
\label{sec:future_thirdparty}

An important research direction is whether stronger \emph{rootless cross-app
intervention} can be achieved for ordinary users. Existing examples suggest
that user-space applications can sometimes observe or even influence another
application without modifying the target app. For example, ARMOUR detects
another application's zero-permission sensor usage entirely from user space,
without root or additional permissions, by exploiting shared behaviors of the
Android sensor framework~\cite{long2025armour}. The work explicitly applied the approach to detecting shake-to-open splash ads. GPS spoofing provides an even
stronger intervention example. After a user selects a mock-location
application, Android allows it to replace location data delivered to other
applications through test location providers~\cite{androidMockLocation}.
These mechanisms
demonstrate that rootless cross-app observation and data mediation are
technically possible under certain system abstractions, although it remains
unclear whether similar design methodologies can be generalized to sensor,
touch, or navigation paths involved in splash advertisements.

Future research could therefore investigate new user-space signals and
system-supported interfaces that enable intervention at common points of the
splash-ad pipeline, while avoiding per-app reverse engineering. Combining
app-agnostic context---such as foreground state, sensor-usage indicators,
interface information, and cross-app transitions---may enable more usable
third-party defenses with lower deployment and maintenance costs than today's
root- and rule-dependent approaches.

\subsection{Incentivize Smartphone Manufacturers for Platform Improvements}
\label{sec:oem_improvement}

In the multi-party ecosystem involving users, application developers,
advertising platforms, and smartphone manufacturers, we believe smartphone
manufacturers are particularly well positioned to drive practical improvements.
Unlike advertising providers whose revenue may depend on advertisement
engagement, manufacturers also compete on overall user experience. In ad previous large-scale user survey~\cite{ma2023ShakeAdPoll},
82\% of respondents in a consumer survey indicated that the availability of
features blocking ``shake-to-open'' redirects could influence their smartphone
purchase decisions. Similar device-level security and
privacy protections, such as anti-peeping notifications and deepfake call
detection, have already been integrated into commercial smartphones
~\cite{oppo2021coloros12,honor2025security,shamsi2025eyehearyou}. User demand could therefore
provide a direct incentive for manufacturers to offer stronger controls over
intrusive splash-ad behavior.

This direction has already begun to emerge. Huawei and OPPO provide
per-application controls over motion or orientation sensors to suppress
shake-triggered advertisement redirects, while HONOR provides system-level
warnings for suspected advertisement-induced cross-application jumps
~\cite{huawei2026orientation,oppoMotionPermissionGuide,honorAdNavigationReminder}. These
mechanisms can protect existing applications without requiring every app or
advertising SDK provider to modify its implementation, although current
solutions remain largely trigger-specific.
Prior work has already demonstrated OS-level fine-grained sensor mediation
~\cite{xu2015semadroid,bolton2023characterizing,long2022side} and privilege separation between advertising
components and host applications~\cite{shekhar2012adsplit}. Compared with
third-party tools that rely on Accessibility, root access, or hooking
frameworks, manufacturer-provided controls could therefore offer a more
deployable path toward systematic protection against intrusive splash ads.

\section{Conclusion}
We studied how splash ads cause unintended app or webpage openings and whether existing ad-blocking tools can help users avoid them. Our analysis and tests show that users face difficulties both setting up these tools and obtaining reliable protection. Some tools failed to stop redirects, and others left apps unable to open or stuck on the ad page. Even tools requiring root access did not consistently work. Smartphone manufacturers and platform providers could reduce this burden by giving users built-in options to disable unwanted ad interactions. The goal is to let users open and use their apps without accidental redirects, while allowing advertising that respects their choices.

\bibliographystyle{ACM-Reference-Format}
\bibliography{references}

\appendix
\onecolumn
\begingroup
\setlength{\emergencystretch}{3em}
\section{Ad Defenses, Deployment Requirements, and Related Mechanisms}
\label{app:defense_inventory}

The main inventory contains 108 advertising countermeasure instances: Appendix~\ref{app:defense_tools} lists 88 splash-related tools, rules, and prototypes; Appendix~\ref{app:defense_settings} lists five user-enabled vendor settings; and Appendix~\ref{app:defense_other} lists 15 instances targeting other ad formats. The resulting 93 splash-related instances include general defenses applicable to in-app advertising, not only tools designed exclusively for splash ads. The other-format instances contribute to the mechanism comparison, but their inclusion does not establish splash-ad coverage.

\textbf{Inclusion and Evidence.} Main-inventory entries have an identifiable ad-related operation supported by source code, a research description, or technical documentation, together with an identified package, service, configuration, or source-build route. Source availability alone is not required, but a general feature claim alone does not establish the operation. Historical versions are retained with their scope and maintenance limits. Appendix~\ref{app:defense_limited} separately records entries with insufficient implementation evidence, incomplete implementations, or unresolved installation routes and required external prerequisites. These entries are excluded from the main-inventory counts, even when their mechanisms are discussed in the main text.

\textbf{Availability and Evaluation.} Mechanism evidence, deployment availability, and tested effectiveness are separate. The operation column describes the mechanism or explicitly attributes a claim; the deployment column describes how users obtain and enable it; the limitations column identifies unresolved conditions. An identified package or source-build route does not establish current device compatibility or successful compilation. Empirical outcomes apply only to the configurations in Table~\ref{tab:defense_empirical}; inclusion here does not imply that a tool passed those tests. Fees and distribution information retain the survey snapshot of September 13, 2026.

\textbf{Counting Unit.} We count tools, rule sets, historical versions, prototypes, and system controls as separate countermeasure instances. For example, GKD and its listed rule subscription are separate entries, as are Cemiuiler and its successor HyperCeiler. Deployment percentages describe the surveyed entries rather than independent products; ecosystems represented by multiple entries contribute more observations. Each instance is counted once in the overall totals and once in each intervention stage it covers. The 22 multistage instances produce 137 stage memberships across the 108 instances.

\textbf{Deployment Labels.} Labels joined by ``+'' within a path are simultaneous requirements; separately listed paths are alternatives. Root / Jailbreak, Runtime Injection, and APK Modification are distinct: injection needs a compatible framework, a module, and a selected target-app scope; APK modification needs a compatible patching workflow, signing, and installation of the modified app. Accessibility and Debugging denote separate authorizations; Shizuku supplies the latter in the listed click and sensor-control paths. VPN denotes VPN authorization, not all network filtering. Overlay and Notification Access denote their respective cross-app permissions. System Setting denotes configuration in an existing operating-system or service interface, rather than an additional permission. Source Build denotes compilation of the defense tool, not merely patching a target APK. Rules inherit the requirements of the listed executor paths. Optional models, certificate trust, and other configuration details are stated separately.

\subsection{Splash-Related Tools, Rules, and Prototypes}
\label{app:defense_tools}
Includes operations explicitly targeting splash ads and general defenses applicable to in-app advertising resources or interfaces.

\begingroup
\fontsize{8.5}{10.5}\selectfont
\setlength{\tabcolsep}{3pt}
\renewcommand{\arraystretch}{1.08}
\begin{longtable}{@{}>{\raggedright\arraybackslash}p{0.14\textwidth}>{\raggedright\arraybackslash}p{0.09\textwidth}>{\raggedright\arraybackslash}p{0.25\textwidth}>{\raggedright\arraybackslash}p{0.25\textwidth}>{\raggedright\arraybackslash}p{0.22\textwidth}@{}}
\caption{Splash-Related Tools, Rules, and Prototypes.}\label{tab:inventory_tools}\\
\toprule
\textbf{Tool / Control} & \textbf{Stage} & \textbf{Operation / Mechanism Evidence} & \textbf{Deployment and Availability} & \textbf{Scope and Limitations} \\
\midrule
\endfirsthead
\multicolumn{5}{l}{\textit{Table \thetable{} (continued)}}\\
\toprule
\textbf{Tool / Control} & \textbf{Stage} & \textbf{Operation / Mechanism Evidence} & \textbf{Deployment and Availability} & \textbf{Scope and Limitations} \\
\midrule
\endhead
\midrule
\multicolumn{5}{r}{\textit{Continued on next page}}\\
\endfoot
\bottomrule
\endlastfoot
Li Tiaotiao rules~\cite{liTiaotiaoCommunityRules} & Presentation / Dismissal & Provide interface-matching and skip-action rules & \textbf{Accessibility}; Rules are executed by Li Tiaotiao and inherit its accessibility and background-execution requirements.\par Li Tiaotiao and compatible rule implementations & Includes non-advertising automation; this appendix includes only advertising rules. \\ \midrule
Ad\-Away~\cite{adaway} & Resource Blocking / Modification & Reject resolution or connections using advertising-domain rules & Resource Blocking / Modification: \textbf{VPN}; Advertising-domain rules\par Resource Blocking / Modification: \textbf{Root / Jailbreak}; Advertising-domain rules.\par Android; local network filtering, or system privileges to modify host mappings & Normal traffic sharing a domain may be blocked; cached ads can still appear. \\ \midrule
Ad\-Blocker Reborn~\cite{adBlockerReborn,adBlockerRebornDelivery} & Resource Blocking / Modification; Initialization / Startup; Presentation / Dismissal & \textbf{Resource Supply:} Intercepts ad requests using URL / hosts rules. \textbf{Initialization and Startup:} Intercepts launches of matching ad Activities. \textbf{Display and Dismissal:} Hides matching ad Views. & \textbf{Root / Jailbreak} + \textbf{Runtime Injection}.\par Android; Xposed \textbf{An APK is included in the repository}; users need not build it from source. & Historical implementation dependent on older interfaces and advertising class names. \\ \midrule
Ad\-Block (Future\-Mind): DNS proxy~\cite{futureMindAdBlock,feeAdblockIos} & Resource Blocking / Modification & A local VPN extension resolves domains in advertising hosts lists to unusable addresses. & \textbf{VPN}; Advertising rules / filtering server and client configuration.\par iOS; VPN extension configuration and imported domain rules. \textbf{Paid.} One-time app purchase includes the local DNS proxy. The same App Store ID is now listed under AdBlock Labs. & Distinct from Safari extensions with the same name and its own Safari module; handles only domains passing through the DNS proxy and cannot distinguish same-origin advertising content. \\ \midrule
Ad\-Close 4.3.7 (verified configuration)~\cite{adClose} & Input / Trigger & Intercepts accelerometer listener registration; the tested request-inspection feature only observes requests. & \textbf{Root / Jailbreak} + \textbf{Runtime Injection}.\par Android; the tested path uses root and Vector 2.2 runtime injection; enable the module and select its host-app scope. & Motion options impose interface-level restrictions without distinguishing advertising from normal motion functions. The SDK initialization interception described in the old README is not implemented in HookLogic in the installed version 4.3.7, and the tested configuration has no network-blocking rules. These two operations are therefore not counted as verified defenses in this version. \\ \midrule
Ad\-Guard DNS: ad-filtering service~\cite{adGuardDnsService,feeAdguardDns} & Resource Blocking / Modification & Filters advertising domains at the resolver, providing a public filtering configuration and customizable private configurations. & \textbf{System Setting}; Ad-filtering DNS address / profile; other client paths follow the respective client's requirements.\par Android / iOS; DNS configuration or a corresponding client. \textbf{Partly paid.} Public ad-filtering DNS and private DNS Starter are free; higher quotas and plans are paid. These are separate from AdGuard app licenses. & Recorded separately from local AdGuard apps as a deployment option; non-filtering configurations do not block ads. Depends on the resolution path and service availability. \\ \midrule
Ad\-Guard for Android~\cite{adguardAndroid,adguardAndroidFeeComparison} & Resource Blocking / Modification & Combine domain filtering with network content filtering & \textbf{VPN}; Advertising rules or filtering server; HTTPS content modification additionally requires certificate trust.\par Android; local VPN or proxy configuration \textbf{Paid.} A license is required for ad blocking in non-browser apps; this function is absent from the free version. & Encrypted-content processing depends on certificate compatibility; features vary by mode and product license. \\ \midrule
Ad\-Guard for i\-OS~\cite{adguardIos,adguardIosNativeAds,feeAdguardIos} & Resource Blocking / Modification & Filter advertising domains accessed by applications and browsers & \textbf{VPN}; Advertising rules / filtering server and client configuration.\par iOS; local filtering or system resolver configuration \textbf{Paid.} Cross-app DNS filtering requires Premium or a purchased AdGuard Pro app; free Safari blocking does not provide this function. & Ads and normal content on the same domain cannot be separated; VPN coexistence conditions differ across modes. \\ \midrule
Ad\-Lock: mobile ad filtering~\cite{adLockMobile} & Resource Blocking / Modification & Uses a local VPN for DNS / HTTP filtering on Android; its cross-app advertising feature on iOS uses DNS filtering. & \textbf{VPN}; Advertising rules or filtering server; HTTPS content modification additionally requires certificate trust.\par Android / iOS; VPN or DNS configuration; HTTPS content inspection additionally requires certificate trust. \textbf{Paid.} The listed cross-app filtering requires a license or subscription; the free iOS Safari module excludes system-wide DNS filtering. & The mechanisms differ between platforms; the free iOS Safari module is not equivalent to cross-app filtering, and DNS mode cannot remove same-origin ads. \\ \midrule
Adobo: advertising patches (loadable by Morphe)~\cite{adoboAdsPatches} & Resource Blocking / Modification; Initialization / Startup & \textbf{Resource Supply:} Statically modifies matching AdMob methods such as load / loadAd to return early. \textbf{Initialization and Startup:} Applies early-return patches to ad startup entry points such as show / showInterstitial. & \textbf{APK Modification}; Install a modified APK, or apply patches, re-sign, and install; this is not equivalent to compiling patch source code.\par Android; a patch manager / build workflow and the modified installation package. \textbf{Compiled patch files are available}; applying them to the target APK and installing it is still required, but compiling patch source is not. & The manager applies patches, while the advertising logic comes from Adobo; failed matching or SDK changes require patch updates, and signatures and app updates introduce compatibility costs. \\ \midrule
Ad\-Popup\-Blocker~\cite{adPopupBlocker} & Initialization / Startup & Terminates forced interstitial or splash entry points according to app rules and invokes adapted game-resumption logic when needed. & \textbf{Root / Jailbreak} + \textbf{Runtime Injection}.\par Android; an environment supporting Modern Xposed API 102. & The inspected version lists only two games; resuming gameplay differs from returning after ad navigation and does not establish general interception of all SDKs. \\ \midrule
Ad\-Skip /\allowbreak{} Android-Touch-Helper~\cite{adSkipTouchHelper} & Presentation / Dismissal & Dismiss ads using keywords, controls, or coordinates & \textbf{Accessibility}; Control rules / recognition configuration and background execution.\par Android; accessibility service & Author notes limited maintenance time and recommends GKD as an alternative. \\ \midrule
Ad\-Skip\-Helper~\cite{adSkipHelper,adSkipHelperDelivery} & Presentation / Dismissal & Scan window nodes and click after matching "Skip" & \textbf{Accessibility} + \textbf{Source Build}; Control rules / recognition configuration and background execution.\par Android; accessibility service \textbf{Source build required:} only source code was found through the inspected public channels; configure the Android SDK / build tools and generate an APK. & Example project; keyword matching may misidentify controls, and internal return values do not establish dismissal. \\ \midrule
ADSkip (Promisin)~\cite{promisinAdSkip} & Presentation / Dismissal & Dismisses splash ads through per-app configurations for control clicks, screen-coordinate clicks, or delayed custom clicks. & \textbf{Accessibility}; Control rules / recognition configuration and background execution.\par Android; accessibility authorization and background execution. & Fixed coordinates and delays depend on layout and loading speed; they do not automatically establish that the target is a genuine close button. \\ \midrule
app2smile ad network rules + Surge /\allowbreak{} Quantumult X~\cite{app2smileAdsense,app2smileSurgeAdsense,app2smileQuantumultAdsense,surgeIosLicenseFees,feeQuantumult} & Resource Blocking / Modification & The corresponding client matches specified advertising endpoints and runs scripts to modify status or result fields in responses from Pangle, Tencent GDT, Kuaishou, and others. & \textbf{VPN}; Rules / scripts; HTTPS content rewriting additionally requires certificate trust; Surge or a compatible client.\par iOS; configurations in the format required by Surge or Quantumult X, and the configuration needed for HTTPS decryption. \textbf{Paid.} Rules / scripts are free; the listed configuration requires a paid Surge Pro / Quantumult X client. Only the chosen compatible client is needed. & The two clients' configurations do not match exactly the same URLs; they depend on endpoints, response structures, and certificate compatibility. Different deployments of the same advertising script are not counted as independent methods. \\ \midrule
Athena: DNS ad filtering~\cite{athenaAdsMode,feeAthenaSource} & Resource Blocking / Modification & DNSRule queries a local rule database and returns a blocking result for requests matching advertising domains. & \textbf{VPN}; Advertising rules / filtering server and client configuration.\par Android; this review confirmed DNS filtering in the VPN packet-processing path. \textbf{Partly paid.} Basic ad filtering is free; custom blocklists require a feature purchase or Premium upgrade. & The project also supports Root / Shizuku modes, but this does not establish equivalent ad-filtering capabilities across all modes; depends on rules and visible DNS queries. \\ \midrule
Auto\-Clicker\-Purifier: advertising features~\cite{autoClickerPurifier,autoClickerPurifierDelivery} & Initialization / Startup; Presentation / Dismissal & \textbf{Initialization and Startup:} Modifies the target app's ad initialization state. \textbf{Display and Dismissal:} Opens the main interface after the splash page's onCreate and calls finish on the original page; calls removeAds to remove banners. & \textbf{Root / Jailbreak} + \textbf{Runtime Injection} + \textbf{Source Build}.\par Android; targets com.zidongdianji; Xposed / LSPosed. \textbf{Source build required:} only source code was found through the inspected public channels; configure the Android SDK / build tools and generate an APK. & Depends on the target app's internal classes and version; its other features are outside this inventory. Splash-page dismissal occurs after onCreate and does not guarantee that the page was not initialized or displayed. \\ \midrule
AWAvenue Ads Rule /\allowbreak{} Qiufeng Advertising Rules~\cite{awaVenueAdsRule} & Resource Blocking / Modification & Provide domain rules for advertising servers and libraries & Resource Blocking / Modification: \textbf{VPN}; Import the compatible domain-rule format into a VPN filtering client\par Resource Blocking / Modification: \textbf{Root / Jailbreak}; Apply the compatible hosts rules through a root-based executor.\par Executed by compatible domain or network filtering tools & A rule set; format conversions derived from the same source share coverage. \\ \midrule
Better\-Heybox (advertising features)~\cite{betterHeybox} & Resource Blocking / Modification; Initialization / Startup & \textbf{Resource Supply:} Removes ad entries from content lists during deserialization. \textbf{Initialization and Startup:} Makes the Xiaoheihe splash-ad function return early. & \textbf{Root / Jailbreak} + \textbf{Runtime Injection}.\par Android; Xposed / LSPosed or a corresponding loading environment & Splash rules depend on fixed obfuscated names; data filtering depends on content tags. \\ \midrule
bili-hook: advertising subset~\cite{biliHook} & Initialization / Startup & Mark Bilibili's regular splash-ad model as invalid & \textbf{Root / Jailbreak} + \textbf{Runtime Injection}.\par Android; libxposed & Inspected documentation specifies client versions and retains birthday splash screens. \\ \midrule
blackmatrix7 startup~\cite{blackmatrixStartup,surgeIosLicenseFees,feeQuantumult,feeLoon} & Resource Blocking / Modification & Remove splash-ad fields and change display duration, dimensions, and expiration & \textbf{VPN}; Rules / scripts; HTTPS content rewriting additionally requires certificate trust; Surge or a compatible client.\par iOS; compatible proxy-rule clients \textbf{Paid.} Rules / scripts are free; the listed configuration requires a paid Surge Pro / Quantumult X / Loon client. Only the chosen compatible client is needed. & Depends on the target application's response format; repository-wide updates do not ensure continued adaptation of this script. \\ \midrule
Block This (historical implementation)~\cite{blockThis} & Resource Blocking / Modification & Send domain queries to the author's configured remote ad-filtering service & \textbf{VPN}; Advertising rules / filtering server and client configuration.\par Android; local VPN configuration & Historical implementation; filtering depends on the remote service, and the client does not maintain a local blocklist. \\ \midrule
Block\-Ads~\cite{blockAdsAndroid} & Resource Blocking / Modification & Filter domains and process encrypted content for selected traffic & \textbf{VPN}; Advertising rules or filtering server; HTTPS content modification additionally requires certificate trust.\par Android; local VPN or privileged proxy & Certificate and protocol compatibility affect connections; some traffic may bypass filtering. \\ \midrule
Blokada 5~\cite{blokadaWebsite,blokadaVersionArchitecture} & Resource Blocking / Modification & Filter advertising traffic locally on the device & \textbf{VPN}; Advertising rules / filtering server and client configuration.\par Android; local VPN \textbf{Partly paid.} Local ad filtering is free; the optional paid Blokada Plus VPN is not required for local filtering. & Listed as an older version on the official website; local rules and device configuration affect filtering. \\ \midrule
Blokada 6 /\allowbreak{} Cloud~\cite{blokadaWebsite,feeBlokadaCloud} & Resource Blocking / Modification & Filter advertising domains through a cloud resolution service & \textbf{System Setting}; Ad-filtering DNS address / profile; other client paths follow the respective client's requirements.\par Android, iOS, and other platforms; filtering-service configuration \textbf{Paid.} Blokada 6 / Cloud requires a filtering-service subscription, unlike the free local mode of Blokada 5. & Depends on the remote service; distinguishes content only by domain. \\ \midrule
CAD viewer patch~\cite{cadViewerAdCleaner} & Initialization / Startup & Remove matching advertising components and change the startup entry to the home screen & \textbf{APK Modification}; Install a modified APK, or apply patches, re-sign, and install; this is not equivalent to compiling patch source code.\par Android; offline package modification and re-signing \textbf{Releases provides a modified APK}; local repackaging is an alternative rather than a prerequisite for using the published package. & Patch for a specific CAD viewer; warm starts, other entry points, and version updates require separate adaptation. \\ \midrule
Cemiuiler: historical advertising features~\cite{cemiuilerMigration} & Initialization / Startup; Presentation / Dismissal & \textbf{Initialization and Startup:} Historical Theme Store rules disable ad support / display eligibility checks. \textbf{Display and Dismissal:} Sets the corresponding ad Views to zero size and hides them. & \textbf{Root / Jailbreak} + \textbf{Runtime Injection}.\par Historical Android / MIUI versions and an Xposed environment. & The repository is archived and recorded as part of the same project's evolution, not as an additional independent technical approach or a currently maintained tool. \\ \midrule
Chaoxing\-Launch\-Ad\-Block~\cite{chaoxingLaunchAdBlock,chaoxingTweakDeployment} & Resource Blocking / Modification; Initialization / Startup & \textbf{Resource Supply:} Blocks Xuexitong splash-ad requests by URL. \textbf{Initialization and Startup:} Makes methods for splash-ad display, window creation, and ad-page entry return early. & \textbf{Root / Jailbreak} + \textbf{Runtime Injection}; Jailbreak and a compatible tweak-loading environment.\par iOS; jailbreaking and Theos / MobileSubstrate loading environment & Targets only Chaoxing Xuexitong; depends on application and system versions. \\ \midrule
Clean\-Splash~\cite{cleanSplashSource,cleanSplashSourceDelivery} & Initialization / Startup; Presentation / Dismissal & \textbf{Initialization and Startup:} Modifies splash-ad enablement / display checks in supported apps. \textbf{Display and Dismissal:} Sets the corresponding Cainiao ad interface elements to GONE. & \textbf{Root / Jailbreak} + \textbf{Runtime Injection} + \textbf{Source Build}.\par Android; Xposed / LSPosed. \textbf{Source build required:} only source code was found through the inspected public channels; configure the Android SDK / build tools and generate an APK. & Documentation targets specified older versions of Cainiao, Bilibili, and QQ; it is not a uniform shake-sensor interceptor. \\ \midrule
Control D: Ads \& Trackers filtering~\cite{controlDAds,feeControlD} & Resource Blocking / Modification & The resolver blocks requests using advertising and tracking domain lists, with different levels of strictness. & \textbf{System Setting}; Ad-filtering DNS address / profile; other client paths follow the respective client's requirements.\par Android / iOS; a DNS endpoint managed by the corresponding Profile. \textbf{Partly paid.} Public Ads \& Trackers resolvers are free; the profile management and custom filtering described in this entry belong to the paid service. & Only this ad-filtering feature is included; stricter rules may block shared legitimate services and require maintaining exceptions. \\ \midrule
Cool\-Apk\-No\-Splash~\cite{coolApkNoSplash,coolApkNoSplashDelivery} & Initialization / Startup & Intercepts a specified Instrumentation launch interface and refuses to open CoolApk's SplashAdActivity. & \textbf{Root / Jailbreak} + \textbf{Runtime Injection}.\par Android; an Xposed / LSPosed module loaded in the CoolApk process. \textbf{CI download:} GitHub Actions provides a prebuilt debug package; downloading generally requires a GitHub account, and artifacts may expire. & Blocks the ad page itself, not the ad's destination app; depends on the component name and launch path. \\ \midrule
Daedalus + Ad\-Away /\allowbreak{} anti-AD rules (historical implementation)~\cite{daedalusAdsSource} & Resource Blocking / Modification & Loads advertising hosts / DNSMasq rules from sources built into the project, matches domains during DNS processing, and returns the addresses specified by the rules. & \textbf{VPN}; Advertising rules / filtering server and client configuration.\par Android; local VPN, with the corresponding advertising rules downloaded and enabled. & The latest confirmed commit on the default branch dates to 2022; compatibility with current Android and availability of old rule URLs were not verified. Domain filtering has difficulty distinguishing normal content from ads on the same domain and does not handle cached resources. \\ \midrule
DNS66~\cite{dnsSixtySix} & Resource Blocking / Modification & Intercept domain queries and match advertising blocklists & \textbf{VPN}; Advertising rules / filtering server and client configuration.\par Android; local VPN & Repository archived; retained as a historical implementation. \\ \midrule
DNSNet~\cite{dnsNet} & Resource Blocking / Modification & Follows the DNS66 approach, matching advertising hosts lists in the DNS request-processing path of a local VPN. & \textbf{VPN}; Advertising rules / filtering server and client configuration.\par Android; VPN authorization and domain rules. & A DNS66 derivative whose maintenance source is recorded separately, rather than a new recognition algorithm; same-origin and cached ads may remain. \\ \midrule
dongqiudi-adblock~\cite{dongqiudiAdblock} & Resource Blocking / Modification & Modifies permissions on Dongqiudi's advertising cache directory and combines hosts, iptables string matching, and IP rules to restrict ad sources. & \textbf{Root / Jailbreak}; Cache permissions, hosts, and network rules.\par Android; Magisk / Root and the relevant network commands. & Cache paths and network rules require continual updates; advertising domains in HTTPS are not directly exposed to plaintext string matching, and rules for shared IP addresses may disrupt normal traffic. \\ \midrule
DTL-X: rmads feature~\cite{dtlxAdsPatches} & Resource Blocking / Modification; Initialization / Startup & \textbf{Resource Supply:} rmads4 replaces matching ad-loading calls with no-ops; ad unit identifiers are also modified. \textbf{Initialization and Startup:} rmads1 removes matching ad Activity declarations; rmads4 also replaces matching show-type calls. & \textbf{APK Modification}; Install a modified APK, or apply patches, re-sign, and install; this is not equivalent to compiling patch source code.\par An Android installation package and an external APK modification / rebuilding toolchain. \textbf{Command-line toolchain:} requires Python and APK unpacking, rebuilding, and signing tools; it is not a defense APK that users simply install. & Coarse-grained and dependent on string patterns; some modes attempt to remove network permission for the entire app and may break functionality or builds. This cannot be described as selective advertising SDK removal. \\ \midrule
Fanqie\-Hook (advertising features)~\cite{fanqieHook} & Initialization / Startup & Modify ad-display decisions and intercept specified splash entry points & \textbf{Root / Jailbreak} + \textbf{Runtime Injection}.\par Android; Xposed / LSPosed or a corresponding loading environment & Targets supported versions of Fanqie Novel and Hongguo Short Drama; retains some user-initiated rewarded ads. \\ \midrule
fmz200/\allowbreak{}wool\_\allowbreak{}scripts: advertising subset~\cite{fmzWoolScripts,fmzMiniAdPluginDependencies,repczToolDependencySnapshot,feeLoon,surgeIosLicenseFees} & Resource Blocking / Modification & Reject ad requests inline and remove or replace advertising fields & \textbf{VPN}; Rules / scripts; HTTPS content rewriting additionally requires certificate trust; Surge or a compatible client.\par iOS; proxy-rule clients supporting the corresponding format \textbf{Paid.} Rules / scripts are free; the listed configuration requires the selected paid Loon / Surge Pro or equivalent client. Only the chosen compatible client is needed. & Some external scripts used by the inspected plugin are missing; inline rules and standalone scripts require separate use. \\ \midrule
Fuck AD~\cite{fuckAd} & Resource Blocking / Modification; Initialization / Startup; Presentation / Dismissal & \textbf{Resource Supply:} Intercepts ad loading and requests. \textbf{Initialization and Startup:} Prevents initialization of supported advertising SDKs. \textbf{Display and Dismissal:} Hides ad Views through code injection or automatically clicks skip controls through an accessibility service. & Resource Blocking / Modification; Initialization / Startup; Presentation / Dismissal: \textbf{Root / Jailbreak} + \textbf{Runtime Injection}\par Presentation / Dismissal: \textbf{Accessibility}; Automatic skipping through accessibility; separate from injection-based hiding\par Presentation / Dismissal: \textbf{Accessibility} + \textbf{Debugging}; Enable Shizuku for click methods using debugging authorization.\par Android; resource supply and initialization interception and View hiding require Xposed / a corresponding code-loading environment; automatic skipping uses an accessibility service, with Shizuku additionally used for some click paths. & Features intervene at different stages and depend on supported advertising libraries and versions. \\ \midrule
Fuck Shake~\cite{fuckShake} & Input / Trigger & Input acquisition: intercepts one sensor listener registration interface & \textbf{Root / Jailbreak} + \textbf{Runtime Injection}.\par Android; Xposed code-loading environment & Does not distinguish advertising sources or sensor types; other interfaces and existing listeners require separate handling. \\ \midrule
Fuck\-Ad\-Jump~\cite{fuckAdJumpSource} & Navigation Control & Intercepts Uri.parse(String) and replaces strings matching JD or Taobao scheme prefixes with empty strings. & \textbf{Root / Jailbreak} + \textbf{Runtime Injection}.\par Android; an Xposed / LSPosed module loaded within its scope. & Uses link prefixes rather than advertising semantics; normal shopping links may also be blocked, while HTTP links and paths bypassing this parsing interface can evade it. \\ \midrule
Fuck\-App (formerly listed as "Zheng Nengliang"): advertising subset~\cite{fuckApp} & Resource Blocking / Modification; Initialization / Startup; Presentation / Dismissal & \textbf{Resource Supply:} Intercepts ad requests on a per-app basis. \textbf{Initialization and Startup:} Intercepts the corresponding advertising SDK or in-app ad entry point. \textbf{Display and Dismissal:} Hides ad interfaces or performs skip clicks. & \textbf{Root / Jailbreak} + \textbf{Runtime Injection}.\par Android; Xposed / LSPosed or a corresponding loading environment & Depends on advertising method names and interface structure; only advertising features are included. \\ \midrule
GKD with specified advertising subscriptions~\cite{gkdAdRules} & Presentation / Dismissal & Match advertising controls with interface selectors and execute actions & \textbf{Accessibility}; Control rules / recognition configuration and background execution.\par Android; accessibility service; some paths can use Shizuku & Forms a defense together with specified advertising subscriptions; other automation rules are excluded. \\ \midrule
Hyper\-Ceiler: advertising features~\cite{hyperCeilerAds} & Resource Blocking / Modification; Initialization / Startup; Presentation / Dismissal & \textbf{Resource Supply:} Wallet rules terminate ad data acquisition and return empty data to the listener. \textbf{Initialization and Startup:} Mi Fitness rules set the splash-ad enablement check to false. \textbf{Display and Dismissal:} Wallet rules hide splash\_container and end splash-ad display. & \textbf{Root / Jailbreak} + \textbf{Runtime Injection}.\par Android / HyperOS; Xposed / LSPosed, subject to the specific module's compatibility. & Only advertising features are listed; system and target-app version changes require adaptation. The current branch's support cannot be attributed to older Cemiuiler versions. \\ \midrule
Invi\-Zible: ad-filtering resolver configuration~\cite{inviZibleAdsMode} & Resource Blocking / Modification & Sends DNS requests to a DNSCrypt resolver that provides ad filtering. & \textbf{VPN}; Advertising rules / filtering server and client configuration.\par Android; selection of an ad-filtering resolver and the corresponding operating mode. & The feature depends on the selected server, and documentation states that the Google Play version does not provide it; DNS encryption itself does not identify ads. \\ \midrule
IVPN: Anti\-Tracker~\cite{ivpnAntiTracker,feeIvpn} & Resource Blocking / Modification & Rejects matching domains through DNS servers maintaining advertising and tracking lists. & \textbf{VPN}; Advertising rules / filtering server and client configuration.\par Android / iOS; the IVPN client and corresponding connection configuration. \textbf{Paid.} AntiTracker is included in paid IVPN plans; no separate ad-blocking add-on purchase is needed. & Although named AntiTracker, the feature explicitly includes ads in official documentation; it remains domain filtering and cannot directly hide ad views. \\ \midrule
Lain-Patches: Disable Ads subset~\cite{lainDisableAds} & Initialization / Startup & Locate advertising functions by method characteristics and return early & \textbf{APK Modification}; Install a modified APK, or apply patches, re-sign, and install; this is not equivalent to compiling patch source code.\par Android; Morphe package-patching workflow \textbf{Compiled patch files are available}; applying them to the target APK and installing it is still required, but compiling patch source is not. & The inspected example targets a specific application and version; only ad-removal patches are included. \\ \midrule
Lin-arm/\allowbreak{}GKD\_\allowbreak{}subscription advertising subset~\cite{linArmGkdRules} & Presentation / Dismissal & Provide control-selection rules for splash and pop-up ads & \textbf{Accessibility}; Rules are executed by GKD and inherit its accessibility and background-execution requirements.\par Executed by GKD & Used together with GKD; rule changes affect target-application support. \\ \midrule
Lockdown Privacy: advertising domain blocking~\cite{lockdownAds,feeLockdown} & Resource Blocking / Modification & Its Firewall feature uses on-device blocking rules for advertising and tracking domains. & \textbf{VPN}; Advertising rules / filtering server and client configuration.\par iOS; Network Extension / local VPN configuration. \textbf{Partly paid.} Basic Firewall has a free path; Advanced Firewall / selected lists and VPN have paid subscriptions. The current store version should not be described as having an entirely free firewall. & Only the explicit ad-blocking feature is recorded; Secure Tunnel is a separate feature. Correspondence between the public repository and the current store version has not been verified. \\ \midrule
madeye/\allowbreak{}ad-skipper~\cite{madeyeAdSkipper} & Presentation / Dismissal & Combine node, text, and image recognition, then click the located target & \textbf{Accessibility}; Control rules / recognition configuration and background execution; screen access and local recognition computation; an additional VLM is optional and is not counted as required for every deployment.\par Android; accessibility screenshots and gestures & Adds recognition overhead; maturity of the optional language-model branch requires separate assessment. \\ \midrule
Mi\-Fitness\-Ad\-Away~\cite{miFitnessAdAway} & Resource Blocking / Modification; Presentation / Dismissal & \textbf{Resource Supply:} Makes splash-ad cache reads return an empty value and filters page ad data. \textbf{Display and Dismissal:} Hides the corresponding ad Views and clears their interface space. & \textbf{Root / Jailbreak} + \textbf{Runtime Injection}.\par Android; system administration privileges and LSPosed / libxposed & Targets Xiaomi fitness and health applications; normal health functions require separate checks. \\ \midrule
Min\-Min\-Guard~\cite{minMinGuard} & Resource Blocking / Modification; Initialization / Startup; Presentation / Dismissal & \textbf{Resource Supply:} GoogleGms intercepts loadAd / loadAds. \textbf{Initialization and Startup:} Intercepts ad startup entry points such as InterstitialAd.show. \textbf{Display and Dismissal:} Sets ad View height to zero to clear occupied space. & \textbf{Root / Jailbreak} + \textbf{Runtime Injection}.\par Android; Xposed & Archived; inspected examples cover banner and interstitial ads. \\ \midrule
mirad-tech/\allowbreak{}Skip~\cite{skipMirad} & Presentation / Dismissal & Locate nodes by rules, recheck targets, then click or issue gestures & \textbf{Accessibility}; Control rules / recognition configuration and background execution.\par Android 9+; accessibility service & Rechecks and cooldown policies are enforced by the tool; still depends on nodes and timing. \\ \midrule
Mullvad: advertising DNS filtering~\cite{mullvadAdsDns,mullvadDnsClosure} & Resource Blocking / Modification & Filters DNS using advertising domain lists, distinguishing DNS within the VPN from public encrypted DNS. & \textbf{System Setting}; Ad-filtering DNS address / profile; other client paths follow the respective client's requirements.\par Android / iOS; the corresponding VPN or DNS configuration. \textbf{Partly paid.} Public ad-filtering DNS is free, whereas filtering inside the VPN requires the paid VPN. Public DNS is scheduled to close on November 2, 2026; the two paths are distinguished. & Public encrypted DNS is announced to shut down on 2026-11-02; this announcement does not imply that DNS within the VPN will also shut down. Alternative resolvers do not automatically provide the same advertising lists. \\ \midrule
Net\-Guard: hosts-based ad filtering in non-Play builds~\cite{netGuardAdsMode,feeNetguard} & Resource Blocking / Modification & Non-Google Play distributions provide hosts-based advertising domain filtering and process the corresponding requests through a local VPN. & \textbf{VPN}; Advertising rules / filtering server and client configuration.\par Android; a build supporting this feature, VPN authorization, and an advertising hosts list. \textbf{Partly paid.} Hosts-based ad filtering in the non-Play build is free; selected advanced features such as logging and traffic analysis require Pro and are not prerequisites for basic ad filtering. & Only the officially documented ad-filtering feature is included, not the general firewall; the Google Play version does not provide this feature. \\ \midrule
Next\-DNS: ad filtering~\cite{nextDnsAds,feeNextDns} & Resource Blocking / Modification & Refuses to resolve matching domains at the remote DNS resolution stage according to selected advertising lists. & \textbf{System Setting}; Ad-filtering DNS address / profile; other client paths follow the respective client's requirements.\par Android / iOS; system DNS settings, a configuration profile, or a client. \textbf{Partly paid.} The first 300,000 queries per month are free; above this quota DNS resolution continues without filtering. Higher usage requires a paid plan. & Depends on the service, lists, and whether queries pass through the resolver; cannot filter specific ad elements within the same domain. \\ \midrule
Nord\-VPN: mobile ad filtering~\cite{nordMobileAds,feeNordAndroid} & Resource Blocking / Modification & Blocks advertising domains on mobile through the DNS service used by the VPN. & \textbf{VPN}; Advertising rules / filtering server and client configuration.\par Android / iOS; a subscription plan supporting the feature and a VPN connection. \textbf{Paid.} Mobile DNS ad filtering is provided through the paid NordVPN service; desktop feature coverage does not apply. & The current official name is Scam and phishing protection, formerly Threat Protection; download scanning or content inspection in the desktop Pro version should not be attributed to mobile versions. \\ \midrule
No\-Shaking\-AD /\allowbreak{} Bu Xu Tiaozhuan~\cite{noShakingAd,noShakingServiceRecovery,noShakingServiceBinding} & Input / Trigger & Input acquisition: restricts the global sensor mode after a foreground-app change and schedules re-enabling after five seconds & \textbf{Accessibility} + \textbf{Debugging}; Enable Shizuku, accessibility, and background execution.\par Android; Shizuku and accessibility service & Affects other applications; preserving prior state and restoring it after abnormal exits have limitations. \\ \midrule
obaby/\allowbreak{}skip\_\allowbreak{}ads\_\allowbreak{}android~\cite{skipAdsObaby} & Presentation / Dismissal & Capture screenshots, locate skip buttons with object detection, and click their coordinates & \textbf{Accessibility}; Control rules / recognition configuration and background execution; screen access and local recognition computation.\par Android; accessibility and screen-capture authorization & Image recognition and page changes may cause misalignment; targets need rechecking before action. \\ \midrule
One Patch: touch blocking for ad regions~\cite{onePatchTouch} & Input / Trigger & After a user selects a rectangular region, an overlay consumes touch events within it. & Input / Trigger: \textbf{Overlay}; User-selected region\par Input / Trigger: \textbf{Accessibility}; User-selected region; an alternative to the regular overlay path.\par Android; floating-window or accessibility-overlay permission. & The official description explicitly lists use for ad regions, but provides no automatic ad recognition; normal buttons are also blocked, while touches outside the region and motion triggers remain. \\ \midrule
personal\-DNSfilter~\cite{personalDnsFilter} & Resource Blocking / Modification & Locally match preconfigured advertising-domain lists & \textbf{VPN}; Advertising rules / filtering server and client configuration.\par Android / Java; platform-specific configuration & Limited for shared-domain content and requests that bypass the resolver. \\ \midrule
Private Internet Access: MACE~\cite{piaMace,feePiaMace} & Resource Blocking / Modification & MACE on Android filters advertising domains through DNS; current official documentation associates the iOS control with Safari content blocking. & \textbf{VPN}; Advertising rules / filtering server and client configuration.\par Android; a client / distribution supporting MACE and the service configuration. \textbf{Paid.} MACE is provided with a PIA subscription and requires a distribution that supports the feature. & Android DNS capabilities cannot be directly described as a defense against native-app ads on iOS; installation channels and platforms must be confirmed separately. \\ \midrule
Proton VPN: Net\-Shield ad filtering~\cite{protonNetShield} & Resource Blocking / Modification & Rejects matching domains when the advertising and tracking blocking level is selected on the DNS service used by the VPN. & \textbf{VPN}; Advertising rules / filtering server and client configuration.\par Android / iOS; a supporting subscription plan, client, and VPN connection. \textbf{Paid.} NetShield requires VPN Plus or another paid Proton VPN plan; the free VPN plan excludes it. & The malware-only level is not equivalent to ad filtering; paths that bypass the corresponding DNS or use Tor do not receive the same protection. \\ \midrule
Pure\-NGA: advertising subset~\cite{pureNGA} & Resource Blocking / Modification; Initialization / Startup; Presentation / Dismissal & \textbf{Resource Supply:} Intercepts SDK ad-loading methods such as load*Ad. \textbf{Initialization and Startup:} Prevents the corresponding SDK initialization and ad startup callbacks. \textbf{Display and Dismissal:} Enters the home page in LoadingActivity foreground callbacks and hides the corresponding ad / recommendation Views. & Resource Blocking / Modification; Initialization / Startup; Presentation / Dismissal: \textbf{Root / Jailbreak} + \textbf{Runtime Injection}; Module and scope\par Resource Blocking / Modification; Initialization / Startup; Presentation / Dismissal: \textbf{APK Modification}; Modified installation package; an alternative to the injection path.\par Android; Xposed / LSPosed or package modification & Depends on NGA's internal fields and startup flow; normal initialization must be preserved. \\ \midrule
Pure\-Skip~\cite{pureSkipSource,pureSkipSourceDelivery} & Presentation / Dismissal & Finds and clicks close buttons during the startup window according to target-app and control rules. & \textbf{Accessibility} + \textbf{Source Build}; Control rules / recognition configuration and background execution.\par Android; an accessibility service. \textbf{Source build required:} only source code was found through the inspected public channels; configure the Android SDK / build tools and generate an APK. & The inspected implementation primarily uses rule-based recognition; coverage changes with built-in adaptations and should not be described as OCR or model-based recognition. \\ \midrule
Qzxy\-Ad\-Block~\cite{qzxyAdBlock,qzxyAdBlockDelivery} & Presentation / Dismissal; Input / Trigger & \textbf{Display and Dismissal:} Exits the Quzhi Campus ad page and removes specified ad Views. \textbf{Input and Evaluation:} Intercepts accelerometer listener registration. & \textbf{Root / Jailbreak} + \textbf{Runtime Injection}.\par Android; all listed interface and sensor operations require Xposed / LSPosed; documentation targets Quzhi Campus 6.5.28. \textbf{An APK is included in the repository}; users need not build it from source. & Spans display and input; interface and sensor rules depend on the specific version, and sensor restrictions may also affect normal functions. Recorded separately from the other TmallCampus project. \\ \midrule
R-Store~\cite{rStore,adCleanerMigration,rStoreMstandConfiguration,surgeIosLicenseFees,feeQuantumult} & Resource Blocking / Modification & Reject ad requests and modify application and mini-program responses & \textbf{VPN}; Rules / scripts; HTTPS content rewriting additionally requires certificate trust; Surge or a compatible client.\par iOS; compatible proxy-rule clients \textbf{Paid.} Rules / scripts are free; the listed configuration requires a paid Surge Pro / Quantumult X client. Only the chosen compatible client is needed. & Successor to Ad-Cleaner; each target's rules depend on specific addresses and scripts. \\ \midrule
Rethink DNS + Firewall: advertising domain rules~\cite{rethinkAdsMode,feeRethink} & Resource Blocking / Modification & Blocks DNS requests entering the filtering path according to a selected advertising domain blocklist. & \textbf{VPN}; Advertising rules / filtering server and client configuration.\par Android; local VPN and the corresponding DNS / rule configuration. \textbf{Partly paid.} The base app and advertising-domain filtering are free; the optional RPN VPN subscription is paid and is not required for basic filtering. & Only the explicit advertising blocklist feature is included; ordinary per-app network blocking is excluded. Depends on the selected lists and DNS path. \\ \midrule
Re\-Weibo: Weibo Lite advertising subset~\cite{reWeibo} & Resource Blocking / Modification; Initialization / Startup & \textbf{Resource Supply:} Filters feed ad data in Weibo Lite. \textbf{Initialization and Startup:} Modifies the startup branch to enter the main interface directly. & \textbf{Root / Jailbreak} + \textbf{Runtime Injection}.\par Android; LSPosed / libxposed & Targets only Weibo Lite; depends on internal methods and versions. \\ \midrule
Shake\-Guard~\cite{shakeGuardSource,shakeGuardSourceDelivery} & Return After Navigation & Matches foreground transitions that have already occurred against source apps, destination apps, and protection periods, then sends one Back action; attempts to reopen the source app when needed. & \textbf{Accessibility} + \textbf{Source Build}; JDK 17, Android SDK 35 / Gradle; source and destination rules and background execution.\par Android; accessibility window events and the global Back action. \textbf{Source build required:} only source code was found through the inspected public channels; configure the Android SDK / build tools and generate an APK. & Acts after navigation and does not read motion data. Tests of the original state-management code reproduced a recovery state that persists after a failed return, and a stale protection-window start after reopening the same app from the launcher. These can suppress later checks or allow new redirects. On-device recovery was not validated. Reopening the source app cannot undo prior network requests or guarantee restoration of the original page. \\ \midrule
SKIP~\cite{skipGuoXiCheng} & Presentation / Dismissal & Click using advertising-control identifiers, pages, and coordinate conditions & \textbf{Accessibility}; Control rules / recognition configuration and background execution.\par Android; accessibility service & Some rules specify screen resolution; layout changes require adaptation. \\ \midrule
Skip\-Ads (Konghuan\-Smart)~\cite{konghuanSkipAds,konghuanSkipAdsDelivery} & Presentation / Dismissal & Matches ad skip controls through an accessibility service and clicks them, with an option to exclude specified apps. & \textbf{Accessibility} + \textbf{Source Build}; Control rules / recognition configuration and background execution.\par Android; an accessibility service. \textbf{Source build required:} only source code was found through the inspected public channels; configure the Android SDK / build tools and generate an APK. & Depends on visible nodes and matching rules; no sensor or final-navigation interception has been established. \\ \midrule
Splash\-Cleaner-Android~\cite{splashCleanerAndroid,splashCleanerAndroidDelivery} & Presentation / Dismissal & Locate dismissal controls using text, identifiers, and structure during a startup scan & \textbf{Accessibility}; Control rules / recognition configuration and background execution.\par Android; accessibility service \textbf{An APK is included in the repository}; users need not build it from source. & Depends on exposed interface information and the scan window; uses node recognition rather than image recognition. \\ \midrule
Splash\-Guard (33lilil)~\cite{splashGuardSource,splashGuardSourceDelivery} & Presentation / Dismissal & Coalesces accessibility window events, finds close buttons using features such as keywords and position, and invokes node clicks. & \textbf{Accessibility} + \textbf{Source Build}; Control rules / recognition configuration and background execution.\par Android; an accessibility service. \textbf{Source build required:} only source code was found through the inspected public channels; configure the Android SDK / build tools and generate an APK. & A small source-code prototype; delayed scans can still encounter stale nodes or deceptive close controls and do not constitute sensor or navigation interception. \\ \midrule
Surfshark: Clean\-Web mobile ad filtering~\cite{surfsharkCleanWeb,feeSurfshark} & Resource Blocking / Modification & Blocks advertising network sources through the VPN app's ad-filtering feature. & \textbf{VPN}; Advertising rules / filtering server and client configuration.\par Android / iOS; the corresponding Surfshark client, subscription plan, and connection settings. \textbf{Paid.} CleanWeb is included in paid Surfshark subscriptions without a separate feature purchase. & Only officially documented in-app ad filtering is included; browser extensions' page-processing capabilities cannot be applied to native apps. Public product descriptions do not establish coverage of each splash-ad category. \\ \midrule
Tap\-Click (formerly ADGO)~\cite{tapClick} & Presentation / Dismissal & Match advertising controls, then click or issue gestures & \textbf{Accessibility}; Control rules / recognition configuration and background execution.\par Android; accessibility service & Former ADGO links redirect to this project; depends on control and coordinate rules. \\ \midrule
The\-Mess\-World: Pangguai Life splash subset~\cite{theMessWorld} & Initialization / Startup & Replace Pangguai Life splash initialization and launch the home screen directly & \textbf{Root / Jailbreak} + \textbf{Runtime Injection}.\par Android; Xposed / LSPosed or a corresponding loading environment & Single-application approach; application hardening and initialization changes affect compatibility. \\ \midrule
Tracker\-Control: advertising-related network filtering~\cite{trackerControlAdsSource} & Resource Blocking / Modification & Processes app connections through a local VPN and blocks matching connections according to user settings, using rules such as domain associations and advertising categories. & \textbf{VPN}; Advertising rules / filtering server and client configuration.\par Android; activation of a version supporting blocking and a local VPN. & Only advertising-related filtering is recorded; blocking advertising tracking connections does not mean removing all ad displays. Coverage depends on domain rules, network paths, and settings; cached and bundled ads may remain. \\ \midrule
Vindroid\-H/\allowbreak{}Skip\-Ads~\cite{skipAdsVindroid} & Presentation / Dismissal & Click using application, window, and keyword rules; fall back to coordinates or back actions & \textbf{Accessibility}; Control rules / recognition configuration and background execution.\par Android; accessibility service & Layout and rule changes affect results; non-advertising automation rules are excluded. \\ \midrule
Windscribe: R.O.B.E.R.T. ad filtering~\cite{windscribeRobert} & Resource Blocking / Modification & Blocks resolution of advertising domains through the DNS-side Ad+Trackers list while connected to Windscribe. & \textbf{VPN}; Advertising rules / filtering server and client configuration.\par Android / iOS; the Windscribe service and corresponding rule configuration. \textbf{Partly paid.} The free plan includes Malware \& Ads; additional rule capacity, lists, and VPN resources are available in paid plans. & Only the named ad-filtering feature is included; requests bypassing this DNS are not covered, and same-origin ads may still appear. \\ \midrule
Wipr 2: Filtr extension~\cite{wiprFiltr,wiprUrlFilterDevelopment,appleNetworkExtensionURLFilters} & Resource Blocking / Modification & Extends ad request filtering beyond Safari to other apps through a system URL filtering extension. & \textbf{System Setting}; Purchase and enable the extension; requires supported iOS / iPadOS versions.\par iOS / iPadOS versions supporting URL Filter; purchase and activation of Filtr. \textbf{Paid.} Requires purchasing Wipr 2 and unlocking the Filtr add-on; cross-app filtering is not part of the base Safari functionality. & The Safari module must be distinguished from the later Filtr release; the store now lists cross-app functionality, so the Safari-only description no longer applies. Coverage depends on URL paths the system can inspect; no splash-ad-specific tests were conducted. \\ \midrule
Xposed-God\-Mode: ad element blocking~\cite{xposedGodMode} & Presentation / Dismissal & Lets users select ad controls, saves rules, and hides the corresponding views in the target app. & \textbf{Root / Jailbreak} + \textbf{Runtime Injection}.\par Android; Xposed / LSPosed. & Uses manual selection rather than automatic ad recognition; reused list controls may cause normal content to be affected. APK export is planned in the documentation and cannot be described as implemented. \\ \midrule
Zero\-Start~\cite{zeroStartSource} & Presentation / Dismissal & First matches accessibility nodes; if matching fails, uses offline text recognition and position rules, then clicks the skip control. & \textbf{Accessibility}; Control rules / recognition configuration and background execution; screen access and local recognition computation.\par Android; an accessibility service; screenshot recognition requires the corresponding screen access capability. & Startup windows, text, and position rules can cause missed detections or incorrect clicks; the shake-prevention effect in the name and description comes from early dismissal, not sensor control. \\ \midrule
Moyu Rewrite: advertising subset~\cite{ddgksApplet,feeQuantumult,surgeIosLicenseFees} & Resource Blocking / Modification & Reject ads by request address or invoke response-modification scripts & \textbf{VPN}; Rules / scripts; HTTPS content rewriting additionally requires certificate trust; Surge or a compatible client.\par iOS; compatible proxy-rule clients \textbf{Paid.} Rules / scripts are free; the listed configuration requires the selected paid Quantumult X / Surge Pro or equivalent client. Only the chosen compatible client is needed. & The inspected mini-program configuration explicitly excludes five-second splash ads; some scripts come from external repositories. \\ \midrule
Li Tiansuo /\allowbreak{} Li\-Tian\-Suo~\cite{liTianSuo} & Initialization / Startup; Presentation / Dismissal; Navigation Control & \textbf{Initialization and Startup:} Intercepts splash display entry points or ad-page launches on a per-app basis. \textbf{Display and Dismissal:} UC rules hide SplashWindow and the corresponding ad Views. \textbf{Navigation:} Intercepts launches of ad landing pages / browser Activities specified in the rule list. & \textbf{Root / Jailbreak} + \textbf{Runtime Injection}.\par Android; libxposed code-loading environment & Covers only listed applications; internal function changes require rule updates. Navigation control is limited to the listed destination components and does not imply interception of arbitrary cross-app navigation. \\ \midrule
Li Tiaotiao (original version and rule ecosystem)~\cite{liTiaotiao,liTiaotiao2023SuspensionReport} & Presentation / Dismissal & Locate and operate skip or dismissal controls using rules & \textbf{Accessibility}; Control rules / recognition configuration and background execution.\par Android; accessibility service & Original version discontinued; original engine, third-party modified packages, and community rules are maintained separately. \\ \midrule
Kuan Jinghua /\allowbreak{} Coolapk Purifier~\cite{coolapkPurifier} & Resource Blocking / Modification; Presentation / Dismissal & \textbf{Resource Supply:} Filters CoolApk's ad entity lists. \textbf{Display and Dismissal:} Finishes the splash page and opens the home page after a splash lifecycle call. & \textbf{Root / Jailbreak} + \textbf{Runtime Injection}.\par Android; libxposed & Page switching occurs after startup; initial adaptation and application versions affect behavior. \\ \midrule
Lei Tiaotiao~\cite{leiTiaoTiao,leiTiaoTiaoImplementation} & Presentation / Dismissal & Match node text or identifiers and click the control or its parent & \textbf{Accessibility}; Control rules / recognition configuration and background execution.\par Android; accessibility service & Consistency between inspected source and same-named shared packages remains unverified; listed by source implementation. \\
\end{longtable}
\endgroup

\subsection{User-Enabled Vendor Advertising Controls}
\label{app:defense_settings}
Users can enable these controls on supported devices and firmware; this does not mean that all Android devices provide the same option.

\begingroup
\fontsize{8.5}{10.5}\selectfont
\setlength{\tabcolsep}{3pt}
\renewcommand{\arraystretch}{1.08}
\begin{longtable}{@{}>{\raggedright\arraybackslash}p{0.14\textwidth}>{\raggedright\arraybackslash}p{0.09\textwidth}>{\raggedright\arraybackslash}p{0.25\textwidth}>{\raggedright\arraybackslash}p{0.25\textwidth}>{\raggedright\arraybackslash}p{0.22\textwidth}@{}}
\caption{User-Enabled Vendor Advertising Controls.}\label{tab:inventory_settings}\\
\toprule
\textbf{Tool / Control} & \textbf{Stage} & \textbf{Operation / Mechanism Evidence} & \textbf{Deployment and Availability} & \textbf{Scope and Limitations} \\
\midrule
\endfirsthead
\multicolumn{5}{l}{\textit{Table \thetable{} (continued)}}\\
\toprule
\textbf{Tool / Control} & \textbf{Stage} & \textbf{Operation / Mechanism Evidence} & \textbf{Deployment and Availability} & \textbf{Scope and Limitations} \\
\midrule
\endhead
\midrule
\multicolumn{5}{r}{\textit{Continued on next page}}\\
\endfoot
\bottomrule
\endlastfoot
OPPO: motion and orientation permission~\cite{oppoMotionPermissionGuide} & Input / Trigger & Input acquisition: denies access to motion or orientation data per app & \textbf{System Setting}; Restricted by device model and OS version.\par User setting listed in the official guide & The guide provides no complete model and firmware compatibility matrix. \\ \midrule
vivo /\allowbreak{} i\-QOO: motion and orientation control~\cite{vivoY500ProMotionControl,vivoOriginOSLaunchMotionRelease} & Input / Trigger & Input acquisition: restricts motion data; some firmware supports blocking only during splash ads & \textbf{System Setting}; Restricted by device model and OS version.\par User setting; models and firmware supporting the option & No uniform duration is officially specified; options require checking for each firmware version. \\ \midrule
Huawei: device orientation permission~\cite{huaweiDeviceOrientationPermission,huaweiShakeAdHelp} & Input / Trigger & Input acquisition: restricts access to orientation data per app & \textbf{System Setting}; Restricted by device model and OS version.\par User setting; corresponding HarmonyOS versions & Normal orientation functions in the application are also restricted. \\ \midrule
HONOR: automatic app navigation reminder~\cite{honorAdNavigationReminder,honorEsg2024AdNavigation} & Navigation Control & Add a reminder for detected accidental ad navigation & \textbf{System Setting}; Restricted by device model and OS version.\par Manufacturer system support and user setting & Scope depends on accidental-activation detection and the navigation being checked. \\ \midrule
HONOR: device sensor permission~\cite{honorMotionPermissionRelease} & Input / Trigger & Input acquisition: restricts motion and orientation access per app & \textbf{System Setting}; Restricted by device model and OS version.\par User setting; HONOR 100 / 100 Pro 8.0.0.150 release notes & Version evidence applies to the listed models; normal motion-based functions may be affected. \\
\end{longtable}
\endgroup

\subsection{Defense Tools for Other Ad Formats}
\label{app:defense_other}
These 15 instances target feed, video, audio, or other non-splash placements. They contribute to the main mechanism and deployment counts, with their advertising scope stated individually.

\begingroup
\fontsize{8.5}{10.5}\selectfont
\setlength{\tabcolsep}{3pt}
\renewcommand{\arraystretch}{1.08}
\begin{longtable}{@{}>{\raggedright\arraybackslash}p{0.14\textwidth}>{\raggedright\arraybackslash}p{0.09\textwidth}>{\raggedright\arraybackslash}p{0.25\textwidth}>{\raggedright\arraybackslash}p{0.25\textwidth}>{\raggedright\arraybackslash}p{0.22\textwidth}@{}}
\caption{Defense Tools for Other Ad Formats.}\label{tab:inventory_other}\\
\toprule
\textbf{Tool / Control} & \textbf{Stage} & \textbf{Operation / Mechanism Evidence} & \textbf{Deployment and Availability} & \textbf{Scope and Limitations} \\
\midrule
\endfirsthead
\multicolumn{5}{l}{\textit{Table \thetable{} (continued)}}\\
\toprule
\textbf{Tool / Control} & \textbf{Stage} & \textbf{Operation / Mechanism Evidence} & \textbf{Deployment and Availability} & \textbf{Scope and Limitations} \\
\midrule
\endhead
\midrule
\multicolumn{5}{r}{\textit{Continued on next page}}\\
\endfoot
\bottomrule
\endlastfoot
Ad Skipper (anar-bastanov)~\cite{anarAdSkipper} & Presentation / Dismissal & Finds and automatically clicks buttons for skippable ads in the official YouTube app. & \textbf{Accessibility}; Control rules / recognition configuration and background execution.\par Android; an accessibility service. & Only for YouTube playback ads, rather than a general splash-ad tool; unskippable ads still require waiting. \\ \midrule
ad-free: audio ad handling~\cite{adFreeAudio,adFreeAudioDelivery} & Presentation / Dismissal & Identifies audio ads using playback state exposed by supported apps, then lowers the volume or plays replacement audio. & \textbf{Notification Access}; Notification / playback-state access and audio control; no accessibility service or injection framework required.\par Android; the corresponding playback-state / notification access and audio control. \textbf{F-Droid provides an installation package}; a source build is not required. & Targets audio apps such as Spotify, rather than splash ads; ads typically continue playing and consuming time. This is not equivalent to removing ads or blocking navigation. \\ \midrule
Ad\-Auto (Hongguo Short Drama ad skipping)~\cite{adAuto} & Presentation / Dismissal & Use advertising terms, countdowns, and playback state to close ads or swipe up & \textbf{Accessibility}; Ad skipping through accessibility; optional auxiliary overlay features are not included in this path.\par Android; the surveyed path uses accessibility. Auxiliary overlay features are outside this path. & Currently focuses on Hongguo Short Drama; swipe actions apply only to the corresponding ad contexts. \\ \midrule
Ad\-Skipper (Gimle\-Larpes)~\cite{gimleAdSkipper,gimleAdSkipperDelivery} & Presentation / Dismissal & Matches YouTube ad button identifiers and clicks automatically; can mute during ads and restore audio afterward. & \textbf{Accessibility} + \textbf{Source Build}; Control rules / recognition configuration and background execution; Android SDK / build tools; muting additionally uses audio control.\par Android; an accessibility service and audio control. \textbf{Source build required:} only source code was found through the inspected public channels; configure the Android SDK / build tools and generate an APK. & Targets YouTube playback ads, rather than splash ads generally; ads continue running while the tool waits for buttons. \\ \midrule
amznkiller~\cite{amznkillerSource} & Presentation / Dismissal & Injects styles into Amazon Shopping's WebView to hide matching sponsored ad elements. & \textbf{Root / Jailbreak} + \textbf{Runtime Injection}.\par Android; an Xposed module loaded in the target app. & Performs interface hiding; ad network requests can still occur. Depends on page structure and selectors and does not cover other native interfaces. \\ \midrule
Discover Feed Filter~\cite{discoverAdsFilter} & Resource Blocking / Modification & Identifies advertising structures in the Google Discover data model and filters the corresponding entries before the list is constructed. & \textbf{Root / Jailbreak} + \textbf{Runtime Injection}.\par Android; Xposed and DexKit for code location. & Targets the Discover feed; changes in advertising structures, obfuscation, and app versions require adaptation. \\ \midrule
Gmail\-Hide\-Ads~\cite{gmailHideAds} & Presentation / Dismissal & Identifies Gmail ad entries when child views are added, makes them invisible, and reduces their height. & \textbf{Root / Jailbreak} + \textbf{Runtime Injection}.\par Android; an Xposed module loaded in Gmail. & Targets native email-list ads, not splash ads; depends on ad view classes, and hiding does not imply request blocking. \\ \midrule
Klick'r /\allowbreak{} Smart Auto\-Clicker: user-configured ad dismissal~\cite{klickrOfficialFeatures,klickrAdCloseReport,feeKlickrFdroid} & Presentation / Dismissal & Users set image or text conditions, and matching triggers actions through an accessibility service; users have reported recognizing and clicking close buttons after game video ads end. & \textbf{Accessibility}; Screen-capture authorization, local image / text matching, user-configured ad-dismissal scenarios, and background execution.\par Android; screen observation, user-configured recognition scenarios, and accessibility service authorization. \textbf{Partly paid.} The current Play version provides all features free but requires watching an ad to start a scenario; a one-time Pro purchase removes the tool's own ads. A free F-Droid build is also available. & An explicit advertising use of a general automation tool, supported by a user report; the user's full scenario file was not obtained, and splash-ad coverage was not verified. Button style changes require rule updates, and recognition and clicking introduce latency. \\ \midrule
Maps\-Ad\-Block~\cite{mapsAdBlock} & Resource Blocking / Modification & Removes advertising entries from data structures after Google Maps responses are parsed. & \textbf{Root / Jailbreak} + \textbf{Runtime Injection}.\par Android; Xposed / LSPosed. & The inspected implementation is restricted to 26.37.00.977222275 and does not install the same hooks on other versions; it is not a general network filter. \\ \midrule
Play Store Adblock~\cite{playStoreAdblock} & Resource Blocking / Modification & Filters sponsored apps and advertising recommendations in the Google Play content-processing path. & \textbf{Root / Jailbreak} + \textbf{Runtime Injection}.\par Android; a libxposed environment, scoped to Google Play. & Primarily targets store search and recommendation ads, rather than splash ads; versions and caches affect results, and modifying the store may also affect its integrity checks. \\ \midrule
QAuxiliary: advertising features~\cite{qAuxiliaryAds} & Resource Blocking / Modification; Presentation / Dismissal & \textbf{Resource Supply:} RemoveCommentAd traverses QQ short-video comment data and removes ad entries. \textbf{Display and Dismissal:} HideQZoneAD / QWalletNoAD hides or removes the corresponding ad Views. & \textbf{Root / Jailbreak} + \textbf{Runtime Injection}.\par Android; module loading in supported QQ / TIM processes. & Only explicit advertising features are counted; different features support different app versions, and the entire utility cannot be treated as a splash-ad defense. \\ \midrule
Threads\-Hide\-Ads~\cite{threadsHideAds} & Resource Blocking / Modification & Locates the feed insertion boundary and filters data entries marked as sponsored content. & \textbf{Root / Jailbreak} + \textbf{Runtime Injection}.\par Android; libxposed and DexKit. & Targets the Threads feed; depends on sponsorship markers and function characteristics, and does not directly restrict touch or motion input. \\ \midrule
Twitter\-Hide\-Ads~\cite{twitterHideAds} & Resource Blocking / Modification; Presentation / Dismissal & \textbf{Resource Supply:} Removes promoted data entries from video lists. \textbf{Display and Dismissal:} Suppresses promoted content at the Compose rendering boundary. & \textbf{Root / Jailbreak} + \textbf{Runtime Injection}.\par Android; a libxposed module loaded in X / Twitter. & Spans data and display; depends on models and rendering entry points and does not establish splash-ad coverage. \\ \midrule
Unclutter\-IG (historical implementation)~\cite{unclutterIg} & Resource Blocking / Modification; Presentation / Dismissal & \textbf{Resource Supply:} Removes sponsored ad data from Stories content lists. \textbf{Display and Dismissal:} Hides Views matching sponsored content after the feed returns ad Views. & \textbf{Root / Jailbreak} + \textbf{Runtime Injection}.\par Android; Xposed. & A historical project whose documented supported versions date from around 2019; not evidence of effectiveness on current Instagram versions. \\ \midrule
XTA-Ad\-Killer~\cite{xtaAdKiller} & Resource Blocking / Modification; Presentation / Dismissal & \textbf{Resource Supply:} Intercepts Super Timetable requests such as loadAd, invokes callbacks with a no-ad result, and clears banner configurations. \textbf{Display and Dismissal:} Sets the corresponding ad Views or containers to GONE. & \textbf{Root / Jailbreak} + \textbf{Runtime Injection}.\par Android; a Modern Xposed module adapted to a specific app. & Not an SDK disabler for arbitrary apps; depends on class-loading timing after decryption, method signatures, and ad callback conventions. \\
\end{longtable}
\endgroup

\subsection{Candidates with Unverified Implementations or Limited Availability}
\label{app:defense_limited}
These 22 entries are excluded from the main-inventory counts. An established mechanism with unresolved access is distinguished from a developer claim or an incomplete implementation; the reason for exclusion is recorded for each entry.

\begingroup
\fontsize{8.5}{10.5}\selectfont
\setlength{\tabcolsep}{3pt}
\renewcommand{\arraystretch}{1.08}
\begin{longtable}{@{}>{\raggedright\arraybackslash}p{0.14\textwidth}>{\raggedright\arraybackslash}p{0.09\textwidth}>{\raggedright\arraybackslash}p{0.25\textwidth}>{\raggedright\arraybackslash}p{0.25\textwidth}>{\raggedright\arraybackslash}p{0.22\textwidth}@{}}
\caption{Candidates with Unverified Implementations or Limited Availability.}\label{tab:inventory_limited}\\
\toprule
\textbf{Tool / Control} & \textbf{Stage} & \textbf{Operation / Mechanism Evidence} & \textbf{Deployment and Availability} & \textbf{Scope and Limitations} \\
\midrule
\endfirsthead
\multicolumn{5}{l}{\textit{Table \thetable{} (continued)}}\\
\toprule
\textbf{Tool / Control} & \textbf{Stage} & \textbf{Operation / Mechanism Evidence} & \textbf{Deployment and Availability} & \textbf{Scope and Limitations} \\
\midrule
\endhead
\midrule
\multicolumn{5}{r}{\textit{Continued on next page}}\\
\endfoot
\bottomrule
\endlastfoot
DNSCloak + Ad\-Guard Home ad filtering~\cite{dnscloakSource,adguardHomeDnscloakSetup,dnscloakSourceDelivery} & Resource Blocking / Modification & The client sends DNS requests over DoH to the configured AdGuard Home service, which filters them using advertising rules. & iOS; DNSCloak network configuration, the server's DNS Stamp, and an AdGuard Home service with ad filtering enabled. \textbf{Historical download unavailable:} the original App Store ID returns no result in the US, China, or UK; the source route needs Xcode, signing, and legacy dependencies, and its buildability has not been verified. & Ad identification is performed by server-side rules, not by DNS encryption itself. The inspected public DNSCloak source was last committed in 2020; this was not used to infer its current distribution status. Apps bypassing this resolution path or using caches may remain unaffected. \\ \midrule
No\-Mo\-Ads (research prototype)~\cite{shuba2018nomoads} & Resource Blocking / Modification & Obtains outbound packets through a local VPN and identifies and blocks ad requests using a classifier trained on network and visible HTTP features. & Android; AntMonitor-based traffic interception; content features depend on decryptability. \textbf{Installation channel unconfirmed:} the paper's project page could not be accessed to verify an APK or a complete build procedure; this entry is excluded from the confirmed source-build count. & A prototype from a 2018 paper; depends on training samples and feature visibility. Certificate pinning makes content features unavailable, and the paper's results do not establish coverage of current splash ads. \\ \midrule
ad\-X /\allowbreak{} ad\-X Launcher~\cite{adxSeven} & Mechanism awaiting verification or limited scope & Vendor describes ad removal in enhanced applications and isolated execution spaces & Android; author-provided runtime environment & Internal modification mechanism is undisclosed; listed as a candidate pending implementation verification. \\ \midrule
Ding Xiaotiao~\cite{adToolCommunityArchive,liTiaotiao2023SuspensionReport} & Mechanism awaiting verification or limited scope & Download records and contemporaneous reports indicate ad-skipping use & Android distribution package; mechanism unverified & Package identity and implementation are unconfirmed; download channels were previously closed. \\ \midrule
Yi Zhi Chan~\cite{adToolCommunityArchive} & Mechanism awaiting verification or limited scope & Ad-skipping application listed in distribution records & Android distribution package; mechanism unverified & Original documentation site is unavailable; correspondence between the package and feature descriptions remains unverified. \\ \midrule
Zhihui Dao~\cite{adToolCommunityArchive} & Mechanism awaiting verification or limited scope & Application in distribution records; advertising operations unverified & Platform and package identity unverified & A renaming relationship with "Chanshi" has not been established, so its implementation description cannot be directly adopted. \\ \midrule
Qing Qidong~\cite{qingQiDong,feeQingLaunch} & Mechanism awaiting verification or limited scope & Author documents accessibility, coordinate clicks, and optional code interception & Android; feature-specific authorization or module loading \textbf{Partly paid.} Distribution materials list a free edition and a paid full edition / activation code. Its implementation-review status is unchanged, and it is excluded from main-table fee counts. & Advertising use is primarily supported by historical reports; current advertising implementation remains unverified. \\ \midrule
Kelee Plugin Center /\allowbreak{} Proxy\-Resource~\cite{keleePlugins,keleePluginCatalog,feeLoon} & Mechanism awaiting verification or limited scope & Provide configuration entries for advertising plugins across applications and mini programs & iOS; Loon; some plugins also require application re-signing \textbf{Paid.} Plugin rules are public; the listed Loon client requires purchase. This entry remains excluded from the main inventory. & Some configurations are inaccessible; the Tencent Video plugin declares discontinued maintenance and version restrictions. \\ \midrule
Ad\-Clear Content Blocker (current product)~\cite{adClearCurrent,adClearContentBlockerIdentity} & Resource Blocking / Modification (specified browsers only) & Filter ads in Samsung Internet / Yandex browsers & Android; specified browser content-filtering interfaces & Current product is browser-only; separate from the historical AdClear native-application filtering product. \\ \midrule
zoffelf/\allowbreak{}skipad (implementation unverified)~\cite{skipadZoffelf,skipadZoffelfDelivery} & Mechanism awaiting verification or limited scope & Author describes node scanning and skip clicks & Android; author states accessibility-service use \textbf{Source only / build needed.} No prebuilt installation package was found; building does not resolve the implementation issues listed here. & Public tree lacks the application implementation and described build workflow; listed as an unverified candidate. \\ \midrule
Yao Ni Ming San Qian: advertising subset (mechanism unverified)~\cite{kwaichingAdSubset} & Mechanism awaiting verification or limited scope & Author lists ad handling in Douyin, Kuaishou, and other applications, including some startup ads & Android; Xposed / LSPosed or a corresponding loading environment & Public distribution tree lacks implementation; specific intervention operations remain unverified. \\ \midrule
Rule Imprison (permissions and state handling awaiting verification)~\cite{ruleImprisonSource,androidApplicationHiddenApi,ruleImprisonSourceDelivery} & Navigation Interception / Confirmation (permissions and state handling awaiting verification) & Monitors when a source app enters the foreground and calls setApplicationHidden to temporarily hide specified shopping apps, restricting destination-app launches; corresponds to ad navigation. & Android; device / profile owner or corresponding delegated privileges, an accessibility service, and compilation after configuring targets in source code. \textbf{Source only / build needed.} No prebuilt installation package was found; building does not resolve the implementation issues listed here. & The initial isHidden value may cause the first hiding call to be skipped; set-active-admin in the documentation does not confer the owner status required by the API. Actual authorization, initial hiding, and restoration need verification. The XML configuration uses typeAllMask, so the dynamic configuration alone does not establish that no events can be received. \\ \midrule
Ad\-Skip\-Tweak (dismissal flow awaiting verification)~\cite{adSkipTweakAudit,adSkipTweakAuditDelivery} & Initialization / Startup Prevention / Presentation Suppression / Dismissal (flow awaiting verification) & Includes advertising-related method replacement, video acceleration, and WebView scripts, affecting ad execution and display. & iOS; a Theos build and a Substrate-style code injection environment. \textbf{CI download:} unexpired build artifacts exist despite the empty Releases page; package contents, signing / injection, and dismissal logic still require checking. & The inspected Makefile compiles only Tweak.xm and does not include AdSkipManager.m, which contains automatic dismissal logic; changing countdown text or accelerating video does not establish that dismissal occurs. General playback acceleration may also affect normal videos. \\ \midrule
Johnny\-Ad\-Block (compatibility awaiting verification)~\cite{johnnyAdblockAudit} & Resource Blocking / Modification / Initialization / Startup Prevention / Presentation Suppression / Dismissal (compatibility awaiting verification) & Installs early-return hooks on methods of specified advertising classes, affecting ad initialization, loading, or display depending on the call. & Android; Xposed / LSPosed. & Actual interception code exists, but the listed class and method names do not establish compatibility with the corresponding SDKs. Target-app versions, methods actually intercepted, and blocking outcomes need to be recorded; exceptions are ignored, so the absence of errors does not demonstrate successful matching. Only advertising features are listed. \\ \midrule
Linked\-In\-Adblocker\-Xposed (unfinished prototype)~\cite{linkedInAdblockerAudit,linkedInAdblockerAuditDelivery} & Unfinished; no ad-removal operation yet & Locates LinkedIn click methods and installs hooks, with the aim of handling in-app ads. & Android; Xposed. \textbf{Source only / build needed.} No prebuilt installation package was found; building does not resolve the implementation issues listed here. & The inspected before / after callbacks are empty and do not yet remove ads. Not counted as an implemented defense, and display-control capabilities are not inferred from the project's goals. \\ \midrule
Re\-Vanced Manager + specific advertising patches (source and versions awaiting verification)~\cite{revancedAdsAudit,revancedAdsPatchLog} & Determined by the specific patch (source code awaiting verification) & Manager executes the package-patching workflow; user logs in the official repository record the Reddit Hide ads patch, with the actual modification point determined by the patch. & Android; the corresponding patch and target-app versions, and installation of the modified package. & Access to the main patch repository returned HTTP 451 during this review, so the specific patch source has not been verified; available logs record patch failure, not successful ad removal. Static modification alone does not place it in ad initialization, and Manager and its patch are not counted twice. \\ \midrule
QQTamer /\allowbreak{} QQ Keeper: historical advertising subset~\cite{qqTamer} & Author-listed functions concern: Resource Blocking / Modification; Navigation Interception / Confirmation (implementation awaiting verification) & Author describes intercepting specified QQ splash-cache use and ad navigation & Android; Xposed / LSPosed or a corresponding loading environment & The public materials do not establish the interception points and supported versions for the claimed cache and navigation functions. Some real-time splash and feed ads are explicitly outside its advertised scope. \\ \midrule
Dasheng Jinghua~\cite{adWars} & Author-listed functions concern: Resource Blocking / Modification; Presentation Suppression / Dismissal (implementation awaiting verification) & Author lists page replacement, view blocking, and request and file rules & Android; Xposed or privileges required by each feature & Distribution materials describe page replacement, interface hiding, and request or file rules, but the available evidence does not link those functions to a working implementation in a specific version. \\ \midrule
Tmall Campus /\allowbreak{} Quzhi Campus ad removal~\cite{tmallCampusAdBlock} & Author-listed functions concern: Resource Blocking / Modification; Initialization / Startup Prevention; Presentation Suppression / Dismissal (implementation awaiting verification) & Author describes direct home-screen entry, hidden ads, and intercepted ad loading & Android; system administration privileges and libxposed & The listed functions are version-specific. The available repository does not establish which calls or interfaces implement home-page entry, interface hiding, or advertising-load interception. \\ \midrule
Adhell3 (historical source)~\cite{adhellThree} & Resource Blocking / Modification & Write advertising-domain and connection rules to Samsung network controls & \textbf{System Setting} + \textbf{Source Build}; Device administrator authorization, a Knox license, and compatible firmware.\par Samsung devices; Knox management, administrator authorization, and a license \textbf{Source build required:} only source code was found through the inspected public channels; configure the Android SDK / build tools and generate an APK. & Historical source documents Samsung Knox rule configuration. Deployment additionally needs administrator authorization and a Knox license; current license availability is unconfirmed. New-device compatibility is also unconfirmed. This entry is excluded because the required external deployment prerequisite is unresolved, not simply because the code is historical. \\ \midrule
Ad\-Skipper (decemberpei)~\cite{adSkipperDecemberpei} & Claimed: presentation / dismissal & The developer describes automatically finding and clicking ad Skip buttons using accessibility interface information. & \textbf{Accessibility}; Control rules / recognition configuration and background execution.\par Android; accessibility authorization and background execution settings. & The store description states automatic ad skipping, but the available evidence does not establish its recognition and intervention mechanism beyond that claim. The implementation is not counted as established. A store listing and a verified defensive operation are separate forms of evidence. \\ \midrule
Bili\-Roaming\-X: advertising patches~\cite{biliRoamingXAds,biliRoamingXAdsDelivery} & Resource Blocking / Modification & Modifies the Bilibili package so that data retrieval for creator-recommended ads returns an empty value; it also clears specified advertising response lists. & \textbf{APK Modification}; ReVanced patching workflow; the prebuilt distribution channel is unavailable, and patch-building and installation channels remain unconfirmed; not counted as a ready-to-install APK.\par Android; the ReVanced patching workflow and installation of the modified app. \textbf{Distribution unavailable at the cited channel:} the official README's prebuilt repository currently returns 404, and the source repository has no release packages; ready-to-install deployment cannot be assumed. & Source code establishes advertising-response patches for Bilibili recommendations. The cited prebuilt repository is unavailable, and a complete usable build-and-install route has not been established. It is therefore retained with availability-limited research and tools, outside the main counts. \\
\end{longtable}
\endgroup

\subsection{Mechanisms Requiring Action by Host-App Developers or Advertising Platforms}
\label{app:defense_cooperative}
Ordinary users cannot independently deploy these interfaces, integration libraries, or platform policies; they are excluded from the main-text method classification and statistics.

\begingroup
\fontsize{8.5}{10.5}\selectfont
\setlength{\tabcolsep}{3pt}
\renewcommand{\arraystretch}{1.08}
\begin{longtable}{@{}>{\raggedright\arraybackslash}p{0.14\textwidth}>{\raggedright\arraybackslash}p{0.09\textwidth}>{\raggedright\arraybackslash}p{0.25\textwidth}>{\raggedright\arraybackslash}p{0.25\textwidth}>{\raggedright\arraybackslash}p{0.22\textwidth}@{}}
\caption{Mechanisms Requiring Action by Host-App Developers or Advertising Platforms.}\label{tab:inventory_cooperative}\\
\toprule
\textbf{Tool / Control} & \textbf{Stage} & \textbf{Operation / Mechanism Evidence} & \textbf{Deployment and Availability} & \textbf{Scope and Limitations} \\
\midrule
\endfirsthead
\multicolumn{5}{l}{\textit{Table \thetable{} (continued)}}\\
\toprule
\textbf{Tool / Control} & \textbf{Stage} & \textbf{Operation / Mechanism Evidence} & \textbf{Deployment and Availability} & \textbf{Scope and Limitations} \\
\midrule
\endhead
\midrule
\multicolumn{5}{r}{\textit{Continued on next page}}\\
\endfoot
\bottomrule
\endlastfoot
Adblock\-Android: developer integration library~\cite{adblockAndroidLibrary} & Resource Blocking / Modification; Presentation / Dismissal & \textbf{Resource Supply:} Checks resource requests against advertising rules in the integrated WebView. \textbf{Display and Dismissal:} Uses ElementHiding to hide matching web ad elements. & Android; integration of the ad-filtering library by the host developer. \textbf{Requires developer integration and a host-app build}; a sample APK does not provide general control over other apps. & Integrated / configured by host-app developers or enabled by advertising platforms; users cannot deploy it themselves. Covers only integrated web containers, rather than allowing users to control other native apps after installation; depends on filtering rules and WebView interfaces. \\ \midrule
libadblockplus-android: Adblock\-Web\-View~\cite{libAdblockplusAndroid} & Resource Blocking / Modification; Presentation / Dismissal & \textbf{Resource Supply:} AdblockWebView filters web ad resource requests. \textbf{Display and Dismissal:} Hides web ad elements. & Android; host integration of the library, subscription configuration, and JavaScript support. \textbf{Requires developer integration and a host-app build}; a sample APK does not provide general control over other apps. & Integrated / configured by host-app developers or enabled by advertising platforms; users cannot deploy it themselves. Spans resource supply and display; does not cover unintegrated native ad interfaces or provide general cross-app control. \\ \midrule
Google Confirmed Click~\cite{googleAdMobConfirmedClick} & Navigation Control & Add confirmation before entering an ad destination & Enabled by Google for selected placements & Integrated / configured by host-app developers or enabled by advertising platforms; users cannot deploy it themselves. Limited to Google Ads demand; specific splash-format coverage requires evaluation. \\ \midrule
Gro\-More Android~\cite{groMoreSplashShakeControl} & Input / Trigger & Trigger evaluation: configures splash-ad shaking through setSplashShakeButton & Developer sets mediation request parameters & Integrated / configured by host-app developers or enabled by advertising platforms; users cannot deploy it themselves. Documentation does not list complete ad-source coverage or default values. \\ \midrule
Sigmob Android~\cite{sigmobSensorInteractionControl} & Input / Trigger & Input acquisition: disables SDK sensor use through setSensorStatus(false) & Developer configuration before ad requests & Integrated / configured by host-app developers or enabled by advertising platforms; users cannot deploy it themselves. Applies to that SDK; its sensor uses also include fraud prevention. \\ \midrule
Sigmob Harmony\-OS~\cite{sigmobSensorInteractionControl} & Input / Trigger & Input acquisition: controls SDK sensor access through isCanUseSensor & Developer supplies privacy-control configuration & Integrated / configured by host-app developers or enabled by advertising platforms; users cannot deploy it themselves. Applies only to the corresponding platform SDK and supported paths. \\ \midrule
Sigmob i\-OS~\cite{sigmobSensorInteractionControl} & Input / Trigger & Trigger evaluation: disables motion interaction components through isCanUseMotionManager & Developer sets the corresponding interface to NO & Integrated / configured by host-app developers or enabled by advertising platforms; users cannot deploy it themselves. The interface contract covers interaction components; other advertising interactions may remain. \\ \midrule
Taku Android~\cite{takuAndroidInteractionControl} & Input / Trigger & Trigger evaluation: configures permitted or prohibited shake, rotation, and swipe interactions & Developer configuration; 6.5.09+; some ad sources have additional version requirements & Integrated / configured by host-app developers or enabled by advertising platforms; users cannot deploy it themselves. Support depends on the ad source, adapter version, and ad format. \\ \midrule
Taku i\-OS~\cite{takuIosShakeConfiguration} & Input / Trigger & Trigger evaluation: configures motion interaction features of ad sources & Developer configuration; 6.4.90+; some sources require newer versions & Integrated / configured by host-app developers or enabled by advertising platforms; users cannot deploy it themselves. Custom sources are unsupported; Kuaishou support is limited to splash and native ads. \\ \midrule
To\-Bid Android~\cite{toBidSensorPrivacyControl} & Input / Trigger & Input acquisition: passes the sensor collection choice to supported ad sources & Developer initialization configuration; 5.0.2+ & Integrated / configured by host-app developers or enabled by advertising platforms; users cannot deploy it themselves. Documentation explicitly describes propagation to Sigmob; other sources require separate verification. \\ \midrule
Tencent GDT Android~\cite{tencentSplashShakeControl} & Input / Trigger & Trigger evaluation: disables splash-ad shaking with shakable="0" & Developer setting per ad request & Integrated / configured by host-app developers or enabled by advertising platforms; users cannot deploy it themselves. Omission or a value of 1 does not block shaking; documentation does not uniquely identify the parameter's first release. \\ \midrule
Kuaishou i\-OS~\cite{kuaishouIosDisableShake} & Input / Trigger & Trigger evaluation: disables splash-ad shaking through disableShake & Developer setting before the request; cited guide is version 3.3.28 & Integrated / configured by host-app developers or enabled by advertising platforms; users cannot deploy it themselves. Not blocked by default; affects only the specified interaction. \\
\end{longtable}
\endgroup

\subsection{General Platform Capabilities Related to Ad Defense}
\label{app:defense_general}
Provided only as technical background; the underlying general capability itself cannot be counted as an ad-specific tool.

\begingroup
\fontsize{8.5}{10.5}\selectfont
\setlength{\tabcolsep}{3pt}
\renewcommand{\arraystretch}{1.08}
\begin{longtable}{@{}>{\raggedright\arraybackslash}p{0.14\textwidth}>{\raggedright\arraybackslash}p{0.09\textwidth}>{\raggedright\arraybackslash}p{0.25\textwidth}>{\raggedright\arraybackslash}p{0.25\textwidth}>{\raggedright\arraybackslash}p{0.22\textwidth}@{}}
\caption{General Platform Capabilities Related to Ad Defense.}\label{tab:inventory_general}\\
\toprule
\textbf{Tool / Control} & \textbf{Stage} & \textbf{Operation / Mechanism Evidence} & \textbf{Deployment and Availability} & \textbf{Scope and Limitations} \\
\midrule
\endfirsthead
\multicolumn{5}{l}{\textit{Table \thetable{} (continued)}}\\
\toprule
\textbf{Tool / Control} & \textbf{Stage} & \textbf{Operation / Mechanism Evidence} & \textbf{Deployment and Availability} & \textbf{Scope and Limitations} \\
\midrule
\endhead
\midrule
\multicolumn{5}{r}{\textit{Continued on next page}}\\
\endfoot
\bottomrule
\endlastfoot
AOSP Sensors off~\cite{aospSensorsOff} & Input / Trigger Restriction (general capability) & Globally stop delivery of the corresponding sensor events & Android developer quick setting; manufacturers may modify it & A general platform capability, not counted as an ad-specific tool. Affects other sensors and normal functionality; not a per-application advertising switch. \\ \midrule
One\-Plus: startup motion permission~\cite{onePlus10TLaunchMotionRelease,onePlus9RLaunchMotionRelease} & Input / Trigger Restriction (general capability) & Deny motion and orientation access during application startup & Official notes for 10T 14.0.0.712 / 9R 14.0.0.603 & A general platform capability, not counted as an ad-specific tool. General startup permission; official documentation does not limit it to advertising control. \\ \midrule
i\-OS browser app launching and link routing~\cite{chromiumIosAppLaunchPolicy,appleUniversalLinkRouting,lineIosSafariLaunchConfirmation} & Navigation Interception / Confirmation (general capability) & Browsers conditionally confirm, allow, or deny external-app launches; the system routes links by association & Browser implementation, system settings, and user choices & A general platform capability, not counted as an ad-specific tool. Browser checks apply only to navigation they handle; link routing itself does not identify ads. \\ \midrule
i\-OS URL Filter~\cite{appleNetworkExtensionURLFilters,appleNetworkExtensionDeployment} & Resource Blocking / Modification (general capability) & The system allows or denies requests using an address dataset & iOS 26+; developers configure entitlements and filtering services & A general platform capability, not counted as an ad-specific tool. Other network stacks must actively participate; Filtr's advertising feature is listed separately under resource supply, and its splash-ad coverage still requires specific testing. \\
\end{longtable}
\endgroup

\subsection{Deployment Requirements and Fees}
\label{app:defense_distribution}

\textbf{Requirement Counts.} Tables~\ref{tab:deployment_distribution} and~\ref{tab:deployment_resources} count the 108 instances in Appendices~\ref{app:defense_tools}--\ref{app:defense_other}. A stage row includes only requirements of the operation at that stage. Each requirement counts an instance once if at least one listed path uses it; alternative-path columns can therefore overlap. The overall row deduplicates instances across stages. A nonzero Root count does not mean that every path for those instances requires root. Network filtering is a mechanism, so it is not treated as a permission column.

AWAvenue rules inherit VPN authorization or root from the respective executor paths. Fuck AD has injection-based hiding, accessibility-based skipping, and Shizuku-assisted click paths; only the last also contributes to Debugging under presentation. AdAuto's auxiliary overlay features and other optional capabilities not established for the listed defensive path are excluded. The counts describe listed paths rather than every possible feature of a multifunction tool.

\begingroup
\fontsize{8.5}{10.5}\selectfont
\setlength{\tabcolsep}{3pt}
\renewcommand{\arraystretch}{1.08}
\begin{longtable}{@{}>{\raggedright\arraybackslash}p{0.24\textwidth}>{\raggedright\arraybackslash}p{0.07\textwidth}>{\raggedright\arraybackslash}p{0.05\textwidth}>{\raggedright\arraybackslash}p{0.095\textwidth}>{\raggedright\arraybackslash}p{0.065\textwidth}>{\raggedright\arraybackslash}p{0.085\textwidth}>{\raggedright\arraybackslash}p{0.08\textwidth}>{\raggedright\arraybackslash}p{0.05\textwidth}>{\raggedright\arraybackslash}p{0.065\textwidth}>{\raggedright\arraybackslash}p{0.07\textwidth}@{}}
\caption{Permissions and Execution Requirements of Listed Paths; Columns May Overlap.}\label{tab:deployment_distribution}\\
\toprule
\textbf{Stage} & \textbf{Instances} & \textbf{VPN} & \textbf{Accessi-\newline bility} & \textbf{Debug.} & \textbf{Root /\newline Jailbreak} & \textbf{Injection} & \textbf{APK} & \textbf{Overlay} & \textbf{Notif.\newline Access} \\
\midrule
\endfirsthead
\multicolumn{10}{l}{\textit{Table \thetable{} (continued)}}\\
\toprule
\textbf{Stage} & \textbf{Instances} & \textbf{VPN} & \textbf{Accessi-\newline bility} & \textbf{Debug.} & \textbf{Root /\newline Jailbreak} & \textbf{Injection} & \textbf{APK} & \textbf{Overlay} & \textbf{Notif.\newline Access} \\
\midrule
\endhead
\midrule
\multicolumn{10}{r}{\textit{Continued on next page}}\\
\endfoot
\bottomrule
\endlastfoot
Resource Blocking / Modification & 58 & 30 & 0 & 0 & 22 & 19 & 3 & 0 & 0 \\ \midrule
Initialization / Startup & 22 & 0 & 0 & 0 & 18 & 18 & 5 & 0 & 0 \\ \midrule
Presentation / Dismissal & 44 & 0 & 24 & 1 & 20 & 20 & 1 & 0 & 1 \\ \midrule
Input / Trigger & 9 & 0 & 2 & 1 & 3 & 3 & 0 & 1 & 0 \\ \midrule
Navigation Control & 3 & 0 & 0 & 0 & 2 & 2 & 0 & 0 & 0 \\ \midrule
Return After Navigation & 1 & 0 & 1 & 0 & 0 & 0 & 0 & 0 & 0 \\ \midrule
\textbf{Overall (Deduplicated)} & \textbf{108} & \textbf{30} & \textbf{27} & \textbf{2} & \textbf{38} & \textbf{35} & \textbf{5} & \textbf{1} & \textbf{1} \\
\end{longtable}
\endgroup

\begingroup
\fontsize{8.5}{10.5}\selectfont
\setlength{\tabcolsep}{3pt}
\renewcommand{\arraystretch}{1.08}
\begin{longtable}{@{}>{\raggedright\arraybackslash}p{0.35\textwidth}>{\raggedright\arraybackslash}p{0.1\textwidth}>{\raggedright\arraybackslash}p{0.16\textwidth}>{\raggedright\arraybackslash}p{0.12\textwidth}>{\raggedright\arraybackslash}p{0.09\textwidth}>{\raggedright\arraybackslash}p{0.1\textwidth}@{}}
\caption{System Configuration, Source Builds, and Fees of Listed Paths.}\label{tab:deployment_resources}\\
\toprule
\textbf{Stage} & \textbf{Instances} & \textbf{System Setting} & \textbf{Source Build} & \textbf{Paid} & \textbf{Partly Paid} \\
\midrule
\endfirsthead
\multicolumn{6}{l}{\textit{Table \thetable{} (continued)}}\\
\toprule
\textbf{Stage} & \textbf{Instances} & \textbf{System Setting} & \textbf{Source Build} & \textbf{Paid} & \textbf{Partly Paid} \\
\midrule
\endhead
\midrule
\multicolumn{6}{r}{\textit{Continued on next page}}\\
\endfoot
\bottomrule
\endlastfoot
Resource Blocking / Modification & 58 & 6 & 0 & 16 & 10 \\ \midrule
Initialization / Startup & 22 & 0 & 2 & 0 & 0 \\ \midrule
Presentation / Dismissal & 44 & 0 & 7 & 0 & 1 \\ \midrule
Input / Trigger & 9 & 4 & 0 & 0 & 0 \\ \midrule
Navigation Control & 3 & 1 & 0 & 0 & 0 \\ \midrule
Return After Navigation & 1 & 0 & 1 & 0 & 0 \\ \midrule
\textbf{Overall (Deduplicated)} & 108 & 11 & 8 & 16 & 11 \\
\end{longtable}
\endgroup

\textbf{Deployment Barriers and Fees.} For 39 of the 108 instances (36.1\%), every listed path requires at least one of root / jailbreak, runtime injection, or APK modification. The other 69 provide at least one path without those operations. This differs from counting any path that uses root: a tool with a VPN alternative belongs to the latter group. No software or service fee was identified for the 39 instances. Among the other 69, 16 require payment for the listed defense and 11 offer free paths alongside paid features or plans. The remaining 42 have no identified software or service fee; this is not a guarantee about future pricing. Client or service charges are inherited by rule configurations, so these counts do not represent 27 independent paid products.

Of the 69 instances with a path without root, injection, or APK modification, 50 use existing apps or services, six require source builds, eight are rule sets, and five are vendor settings. Existing packages can still require VPN or accessibility authorization, rules, certificates, debugging setup, or compatible devices. Fee definitions follow Table~\ref{tab:defense_taxonomy}.

\endgroup

\end{document}